%% file: tlj_secvtx_prd.tex
\documentclass[aps,prd,twocolumn,superscriptaddress,showpacs]{revtex4}

\usepackage{graphicx}
\usepackage{dcolumn}
\usepackage{bm}

\begin{document}

\preprint{FERMILAB-PUB-04-275-E}
\preprint{CDF-7138}

\title{Measurement of the $\bm{t\bar t}$ Production Cross Section
  in $\bm{p\bar p}$ collisions at $\bm{\sqrt{s}=1.96}$ TeV using
  Lepton + Jets Events with Secondary Vertex b-tagging}



\input{Run2Authors_REVTeX4}

\collaboration{CDF Collaboration}
\noaffiliation

\date{\today}

\newcommand{\qqbar}{\ensuremath{q\bar q}}
\newcommand{\Wenu}{\ensuremath{W\rightarrow e \nu}}
\newcommand{\Zee}{\ensuremath{Z\rightarrow e^{+}e}^{-}}
\newcommand{\met}{\ensuremath{E\kern-0.6em\lower-.1ex\hbox{/}_T}}

\begin{abstract}
  We present a measurement of the $t\bar t$~production cross section
  using events with one charged lepton and jets from $p\bar
  p$~collisions at a center-of-mass energy of $1.96\,\textrm{TeV}$.
  In these events, heavy flavor quarks from top quark decay are
  identified with a secondary vertex tagging algorithm.  From
  $162\,\rm{pb}^{-1}$ of data collected by the Collider Detector at
  Fermilab, a total of 48 candidate events are selected, where
  $13.5\pm 1.8$ events are expected from background contributions.  We
  measure a $t\bar t$~production cross section of
  $5.6^{+1.2}_{-1.1}\rm{(stat.)}^{+0.9}_{-0.6}\rm{(syst.)}\,\rm{pb}$.

  \vspace{0.5cm}  
\end{abstract}

\pacs{13.85.Ni, 13.85.Qk, 14.65.Ha}

\maketitle

\input{introduction}

\input{detector}

\input{analysis_overview}

\input{secvtx_studies}
\input{hf_fractions}

\input{backgrounds}

\input{cross_checks}

\input{optimization}
\input{single_tags}

\input{double_tags}
\input{conclusions}

\appendix*
\input{secvtx_appendix}

\begin{acknowledgments}
  We acknowledge many fruitful discussions we have had with
  Michelangelo Mangano.  We also thank the Fermilab staff and the
  technical staffs of the participating institutions for their vital
  contributions. This work was supported by the U.S. Department of
  Energy and National Science Foundation; the Italian Istituto
  Nazionale di Fisica Nucleare; the Ministry of Education, Culture,
  Sports, Science and Technology of Japan; the Natural Sciences and
  Engineering Research Council of Canada; the National Science Council
  of the Republic of China; the Swiss National Science Foundation; the
  A.P. Sloan Foundation; the Bundesministerium f\"ur Bildung und
  Forschung, Germany; the Korean Science and Engineering Foundation
  and the Korean Research Foundation; the Particle Physics and
  Astronomy Research Council and the Royal Society, UK; the Russian
  Foundation for Basic Research; the Comisi\'on Interministerial de
  Ciencia y Tecnolog\'{\i}a, Spain; in part by the European
  Community's Human Potential Programme under contract HPRN-CT-20002,
  Probe for New Physics.
\end{acknowledgments}

\bibliography{tlj_secvtx_prd}

\end{document}

%% file: Run2Authors_REVTeX4.tex
 
\affiliation{Institute of Physics, Academia Sinica, Taipei, Taiwan 11529, Republic of China}  
\affiliation{Argonne National Laboratory, Argonne, Illinois 60439}  
\affiliation{Institut de Fisica d'Altes Energies, Universitat Autonoma de Barcelona, E-08193, Bellaterra (Barcelona), Spain}  
\affiliation{Istituto Nazionale di Fisica Nucleare, University of Bologna, I-40127 Bologna, Italy}  
\affiliation{Brandeis University, Waltham, Massachusetts 02254}  
\affiliation{University of California at Davis, Davis, California  95616}  
\affiliation{University of California at Los Angeles, Los Angeles, California  90024}  
\affiliation{University of California at San Diego, La Jolla, California  92093}  
\affiliation{University of California at Santa Barbara, Santa Barbara, California  93106}  
\affiliation{Instituto de Fisica de Cantabria, CSIC-University of Cantabria, 39005 Santander, Spain}  
\affiliation{Carnegie Mellon University, Pittsburgh, PA  15213}  
\affiliation{Enrico Fermi Institute, University of Chicago, Chicago, Illinois 60637}  
\affiliation{Joint Institute for Nuclear Research, RU-141980 Dubna, Russia} 
\affiliation{Duke University, Durham, North Carolina  27708}  
\affiliation{Fermi National Accelerator Laboratory, Batavia, Illinois 60510}  
\affiliation{University of Florida, Gainesville, Florida  32611}  
\affiliation{Laboratori Nazionali di Frascati, Istituto Nazionale di Fisica Nucleare, I-00044 Frascati, Italy}  
\affiliation{University of Geneva, CH-1211 Geneva 4, Switzerland}  
\affiliation{Glasgow University, Glasgow G12 8QQ, United Kingdom} 
\affiliation{Harvard University, Cambridge, Massachusetts 02138}  
\affiliation{The Helsinki Group: Helsinki Institute of Physics; and Division of High Energy Physics, Department of Physical Sciences, University of Helsinki, FIN-00044, Helsinki, Finland} 
\affiliation{Hiroshima University, Higashi-Hiroshima 724, Japan}  
\affiliation{University of Illinois, Urbana, Illinois 61801}  
\affiliation{The Johns Hopkins University, Baltimore, Maryland 21218}  
\affiliation{Institut f\"{u}r Experimentelle Kernphysik, Universit\"{a}t Karlsruhe, 76128 Karlsruhe, Germany}  
\affiliation{High Energy Accelerator Research Organization (KEK), Tsukuba, Ibaraki 305, Japan}  
\affiliation{Center for High Energy Physics: Kyungpook National University, Taegu 702-701; Seoul National University, Seoul 151-742; and SungKyunKwan University, Suwon 440-746; Korea}  
\affiliation{Ernest Orlando Lawrence Berkeley National Laboratory, Berkeley, California 94720}  
\affiliation{University of Liverpool, Liverpool L69 7ZE, United Kingdom}  
\affiliation{University College London, London WC1E 6BT, United Kingdom}  
\affiliation{Massachusetts Institute of Technology, Cambridge, Massachusetts  02139}  
\affiliation{Institute of Particle Physics: McGill University, Montr\'{e}al, Canada H3A~2T8; and University of Toronto, Toronto, Canada M5S~1A7}  
\affiliation{University of Michigan, Ann Arbor, Michigan 48109}  
\affiliation{Michigan State University, East Lansing, Michigan  48824}  
\affiliation{Institution for Theoretical and Experimental Physics, ITEP, Moscow 117259, Russia}  
\affiliation{University of New Mexico, Albuquerque, New Mexico 87131}  
\affiliation{Northwestern University, Evanston, Illinois  60208}  
\affiliation{The Ohio State University, Columbus, Ohio  43210}  
\affiliation{Okayama University, Okayama 700-8530, Japan} 
\affiliation{Osaka City University, Osaka 588, Japan}  
\affiliation{University of Oxford, Oxford OX1 3RH, United Kingdom}  
\affiliation{University of Padova, Istituto Nazionale di Fisica Nucleare, Sezione di Padova-Trento, I-35131 Padova, Italy}  
\affiliation{University of Pennsylvania, Philadelphia, Pennsylvania 19104}  
\affiliation{Istituto Nazionale di Fisica Nucleare, University and Scuola Normale Superiore of Pisa, I-56100 Pisa, Italy}  
\affiliation{University of Pittsburgh, Pittsburgh, Pennsylvania 15260}  
\affiliation{Purdue University, West Lafayette, Indiana 47907}  
\affiliation{University of Rochester, Rochester, New York 14627}  
\affiliation{The Rockefeller University, New York, New York 10021}  
\affiliation{Istituto Nazionale di Fisica Nucleare, Sezione di Roma 1, University di Roma ``La Sapienza," I-00185 Roma, Italy} 
\affiliation{Rutgers University, Piscataway, New Jersey 08855}  
\affiliation{Texas A\&M University, College Station, Texas 77843}  
\affiliation{Texas Tech University, Lubbock, Texas 79409}  
\affiliation{Istituto Nazionale di Fisica Nucleare, University of Trieste/\ Udine, Italy}  
\affiliation{University of Tsukuba, Tsukuba, Ibaraki 305, Japan}  
\affiliation{Tufts University, Medford, Massachusetts 02155}  
\affiliation{Waseda University, Tokyo 169, Japan}  
\affiliation{Wayne State University, Detroit, Michigan  48201}  
\affiliation{University of Wisconsin, Madison, Wisconsin 53706}  
\affiliation{Yale University, New Haven, Connecticut 06520}  
\author{D.~Acosta}
\affiliation{University of Florida, Gainesville, Florida  32611} 
\author{J.~Adelman}
\affiliation{Enrico Fermi Institute, University of Chicago, Chicago, Illinois 60637} 
\author{T.~Affolder}
\affiliation{University of California at Santa Barbara, Santa Barbara, California  93106} 
\author{T.~Akimoto}
\affiliation{University of Tsukuba, Tsukuba, Ibaraki 305, Japan} 
\author{M.G.~Albrow}
\affiliation{Fermi National Accelerator Laboratory, Batavia, Illinois 60510} 
\author{D.~Ambrose}
\affiliation{University of Pennsylvania, Philadelphia, Pennsylvania 19104} 
\author{S.~Amerio}
\affiliation{University of Padova, Istituto Nazionale di Fisica Nucleare, Sezione di Padova-Trento, I-35131 Padova, Italy} 
\author{D.~Amidei}
\affiliation{University of Michigan, Ann Arbor, Michigan 48109} 
\author{A.~Anastassov}
\affiliation{Rutgers University, Piscataway, New Jersey 08855} 
\author{K.~Anikeev}
\affiliation{Massachusetts Institute of Technology, Cambridge, Massachusetts  02139} 
% ...

%% file: introduction.tex
\section{\label{sec:introduction}Introduction}

The top quark is pair-produced in $p\bar p$ collisions through
quark-antiquark annihilation and gluon-gluon fusion. The measurement
of the $t\bar t$ cross section tests the QCD calculations for the pair
production of a massive colored triplet. These calculations have been
performed in perturbation theory to next-to-leading
order~\cite{cacciari,kidonakis}.  Recent work on corrections for soft
gluon emission show that their effect on the cross section is small,
and that they reduce the theoretical uncertainty arising from the
choice of renormalization and factorization scales to less than 5\%
over the expected range of top masses and parton distribution
functions (PDFs).  The leading theoretical uncertainties are in the
PDFs, arising mostly from the understanding of the gluon
distributions at large parton $x$. The total theoretical uncertainty
is approximately 15\%~\cite{cacciari}. At $\sqrt{s}$=1.96 TeV, the
predicted $t\bar t$~production cross section is $\sigma_{t\bar t} =
6.7^{+0.7}_{-0.9}$ pb at $m_t =
175\,\mathrm{GeV}/c^2$~\cite{cacciari}.  For every
$1\,\mathrm{GeV}/c^2$ increase in the top mass over the interval $170
< m_t < 190\,\mathrm{GeV}/c^2$, the $t\bar t$ cross section decreases by
0.2 pb.

The Standard Model top quark decays to a $W$ boson and a $b$ quark
almost 100\% of the time. Top quark pair production thus gives rise to
two $W$ bosons and two ``$b$~jets'' from $b$ quark fragmentation.
When exactly one $W$ decays leptonically, the $t\bar t$~event
typically contains a high transverse momentum lepton, missing
transverse energy from the undetected neutrino, and four high
transverse momentum jets, two of which originate from $b$ quarks.
This mode is labelled ``$W$ plus jets'' or ``lepton plus jets.''
Since the final state branching ratio is directly related to the $W$
branching ratios, the $t\bar t$ rate into a particular final state
measures both the production and decay properties of the top quark. An
unexpected result could thus indicate either a non-standard source of
top-like events, or a modification of the top decay branching ratios.

The $p\bar p$~collisions for this measurement of $t\bar t$ production
were produced during Run II of the Fermilab Tevatron. The data were
recorded at CDF II, a general purpose detector which
combines charged particle tracking, sampling calorimeters,
and fine-grained muon detection. Isolating the lepton plus jets decay
mode of the top quark builds on the detailed understanding of
inclusive leptonic $W$ boson decays in CDF II~\cite{ewkpaper}.  The
$t\bar t$ signature is mimicked by processes in which a $W$ boson is
produced in association with several hadronic jets with large
transverse momentum.  To separate the $t\bar t$ events from this
background we use precision silicon tracking to $b$-tag jets
containing a secondary vertex from a $b$ hadron decay.  Background
contributions from fake $W$s, mis-identified secondary vertices and
heavy flavor production processes such as $Wb\bar b$ are estimated
using a combination of Monte Carlo calculations and independent
measurements in control data samples.  An excess in the number of events
which contain a lepton, missing energy, and three or more jets with at least
one $b$-tag is the signal of $t\bar t$ production and is used to
measure the production cross section $\sigma_{t\bar t}$. The dataset
defined by this analysis forms the basis for other measurements of top
quark properties, such as the top quark mass and the helicity of $W$
bosons produced in top decays.

This measurement builds on the $b$-tagging techniques employed by CDF 
at the Tevatron Run~I. Then, at $\sqrt{s}$=1.8 TeV, a similar analysis of
lepton+jets events with $b$-tags gave a $t\bar t$ cross section of
$\sigma_{t\bar t} = 5.1 \pm 1.5$ pb~\cite{run1prd}, compared to an
expected value of $\sigma_{t\bar t} = 5.2^{+0.5}_{-0.7}$ pb at
$m_{\mathrm t} = 175\,\mathrm{GeV}/c^2$~\cite{cacciari}.  Here, using
a larger dataset collected at higher center-of-mass energy as well as 
improved Monte Carlo tools and detector simulations, we have
re-analyzed the heavy flavor fraction in $W$ events and improved our
understanding of $b$-tagging efficiencies, including the contribution
of material interactions to fake $b$ tags. In addition, the
significance of the measurement is optimized by requiring a large
scalar sum of the transverse energies of all objects in the event
($H_T$), which improves the rejection of background events.  

Our analysis complements other recent $t\bar t$ cross section
determinations at CDF II using dilepton events~\cite{dilepton} or using
lepton plus jets events with $b$-tags and a kinematically derived
estimate of the $b$-tagged backgrounds~\cite{chicago}. The work of Reference  
~\cite{chicago} is particularly relevant to the measurement described here, 
in that it uses the same b-tagged event sample, but calculates the 
backgrounds by appealing to a data control sample available only in 
the $W$ plus jets selection. Our technique for background estimation is 
significantly more general, and this paper establishes the ability to 
use b-tagging in many other kinds of measurements at CDF in the future. 
We comment further on this matter at the end of Section III. 

The organization of this paper is as follows.  Sec.~\ref{sec:detector}
reviews the detector systems and event reconstruction techniques
relevant to this measurement.  The trigger and sample selections are
described in Sec.~\ref{sec:data}. The $b$-tagging algorithm, its
efficiency for tagging $b$ jets, and the understanding of its fake
rate are discussed in Sec.~\ref{sec:secvtx}.  The means for estimating
backgrounds from processes which produce a $W$ in association
with heavy flavor are described in Sec.~\ref{sec:hffracs}.  In
Sec.~\ref{sec:m2bkg} our understanding of mistags and backgrounds is
applied to collate a comprehensive estimate of all tagged
contributions to the lepton + jets sample, and this estimate is
compared with the data.  A cross-check of the background estimation,
using the $Z$ + jets sample, is presented in Sec.~\ref{sec:checks}.
In Sec.~\ref{sec:optimization} an optimization using the total
transverse energy in the event to improve the cross section
measurement uncertainty is described, along with the acceptance
associated with this event selection.  The $t\bar t$ production cross
section measured in events with at least one $b$-tagged jet is
presented in Sec.~\ref{sec:singletags}; the result in events with at
least two $b$-tagged jets is presented in Sec.~\ref{sec:dbtag}.  The
final results are summarized in Sec.~\ref{sec:conclusions}.

%% file: detector.tex
\section{\label{sec:detector}Event Detection and Reconstruction}

The CDF II detector is described using a cylindrical coordinate system
with the $z$ coordinate along the proton direction, the azimuthal
angle $\phi$, and the polar angle $\theta$ usually expressed through
the pseudorapidity $\eta = -\ln(\tan(\theta/2))$.  The rectangular
coordinates $x$ and $y$ point radially outward and vertically upward
from the Tevatron ring, respectively.  The detector is approximately
symmetric in $\eta$ and $\phi$.

\subsection{Charged Particle Tracking}

Drift cell and silicon microstrip systems provide charged particle
tracking information in the region $|\eta|\leq 1.0$ and $|\eta|\leq
2.0$, respectively.  The tracking systems are contained in a
3.2 m diameter, 5 m long superconducting solenoid which produces a
$1.4\,\mathrm{T}$ magnetic field aligned coaxially with the $p\bar p$
beams, allowing measurement of charged particle momentum transverse to
the beamline ($p_T$).

The Central Outer Tracker (COT) is a 3.1 m long open cell drift
chamber which performs 96 track measurements in the region
between 0.40 and 1.37 m from the beam axis~\cite{cot}.  
Sense wires are arranged in 8 alternating axial and
$\pm2^\circ$ stereo ``superlayers" with 12 wires each. The position
resolution of a single drift time measurement is approximately
$140\,\mu{\rm m}$.

Charged particle trajectories are found first as a series of
approximate line segments in the individual axial superlayers. Two
complementary algorithms are used to associate segments lying on
a common circle, and the results are merged to yield a final set of
axial tracks. Track segments in the stereo superlayers are associated with
axial track segments to reconstruct tracks in three dimensions.
COT tracks used in this analysis are
required to have at least 3 axial and 3 stereo superlayers with 7 hits
per superlayer.

The efficiency for finding isolated high momentum tracks is measured 
using electrons from $W \rightarrow e^{\pm} \nu$ which are identified 
in the central region $|\eta| \leq 1.1$ using only the calorimetric information 
for the electron shower and the missing transverse energy (see below). 
In these events, the efficiency for finding the electron track 
is found to be $99.93^{+0.07}_{-0.35}\%$, and this is 
typical for high momentum isolated tracks from either
electronic or muonic W decays which are contained in the COT.   
For high-momentum tracks, the transverse momentum resolution is found to be
$\delta p_T/p_T \approx 0.1\% \cdot p_T$(GeV), the track position 
resolution at the origin is $\delta z \approx
0.5\,\mathrm{cm}$ in the direction along the beamline and
the resolution on the track impact parameter, or distance from the beamline at
the track's closest approach in the transverse plane, 
is $\delta d_0 \approx 350\mu{\rm m}$.

A road-based hardware pattern recognition algorithm runs online in the
eXtremely Fast Tracker (XFT) to provide track information for
triggering~\cite{xft}.  Drift times partitioned into two time bins are
used to find the axial segments which are matched in their positions
and slopes.  An ``XFT track" is one which has four matching axial
segments on a trajectory. The XFT efficiency is measured in a set of 
well measured COT tracks  which pass through 
all 4 axial superlayers. The XFT is found to have an average
efficiency of $96.7 \pm 0.1\%$ for charged particles with momenta
greater than $25\,\mathrm{GeV}/c$.

Inside the inner radius of the COT, a five layer double-sided silicon
microstrip detector (SVX) covers the region between 2.5 to 11 cm from
the beam axis~\cite{silicon}. Three separate SVX barrel modules are
juxtaposed along the beamline to cover a length of 96 cm,
approximately 90\% of the luminous beam intersection region.  Three of
the five layers combine an $r-\phi$ measurement on one side and a
$90^\circ$ stereo measurement on the other, and the remaining two
layers combine $r-\phi$ with small angle stereo at $\pm1.2^\circ$.
The typical silicon hit resolution is $11\,\mu\mathrm{m}$.  Additional
Intermediate Silicon Layers (ISL) at radii between 19 and 30 cm in the
central region link tracks in the COT to hits in SVX.

Silicon hit information is added to reconstructed COT tracks using a
progressive ``Outside-In'' (OI) tracking algorithm. COT tracks are
extrapolated into the silicon detector, associated silicon hits are
found, and the track is refit with the added information of the
silicon measurements. The initial track parameters provide a width for
a search road in a given layer.  Then, for each candidate hit in that
layer, the track is refit and used to define the search road into the
next layer. The stepwise addition of the precision SVX information at
each layer progressively reduces the size of the search road, while
also properly accounting for the additional uncertainty due to
multiple scattering in each layer. The search uses the two best
candidate hits in each layer to generate a small tree of final track
candidates, from which the tracks with the best $\chi^2$ are selected.
The efficiency for associating at least three silicon hits with an
isolated COT track is $91 \pm 1\%$.  The extrapolated
impact parameter resolution for high momentum OI tracks is $30\,\mu
\mathrm{m}$, including the uncertainty in the beam position.

\subsection{Calorimetry for Electrons and Jets}

Outside of the tracking systems and the solenoid, segmented
calorimeters with projective geometry are used to reconstruct
electromagnetic (EM) showers and jets~\cite{cem,had,plug}. 
The EM and hadronic calorimeters are
lead-scintillator and iron-scintillator sampling devices, respectively.
The calorimeter is segmented into ``towers'', each covering a 
small range of pseudo-rapidity and azimuth; the full array covers
$2 \pi$ azimuth over the pseudo-rapidity range $|\eta|<3.6$.   
The transverse energy $E_T = E\sin{\theta}$ is measured in each
calorimeter tower, where the polar angle is calculated using the measured
$z$ position of the event vertex.  Proportional and scintillating
strip detectors measure the transverse profile of EM showers at a
depth corresponding to the shower maximum.

High momentum jets, photons, and electrons leave isolated energy
deposits in small contiguous groups of calorimeter towers which can be
identified and summed together into an energy ``cluster.''  For the
purpose of triggering, online processors organize the calorimeter
tower information into separate lists of clusters for the
electromagnetic compartments alone and for the electromagnetic and
hadronic compartments combined.  Electrons are identified in the
central electromagnetic calorimeter (CEM) as isolated, mostly
electromagnetic clusters which match with an XFT track, in the
pseudorapidity range $|\eta|<1.1$.

The electron transverse energy is reconstructed from the
electromagnetic cluster with a precision $\sigma(E_T)/E_T =
13.5\%/\sqrt{E_T/(\mathrm{GeV})} \oplus 2\%$~\cite{emcal}.  Jets are
identified as a group of electromagnetic and hadronic calorimeter
clusters which fall within a cone of radius $\Delta{R}=\sqrt{\Delta
  \phi^2 + \Delta \eta^2} \leq 0.4$~\cite{jetclu}.  Jet energies are
corrected for calorimeter non-linearity, losses in the gaps between
towers~\cite{hadroncal}, and multiple primary interactions. The jet 
energy resolution is approximately
$(0.1(E_T/(\mathrm{GeV})) + 1.0)$ GeV~\cite{jetres}.

\subsection{Muon Detection and Reconstruction}

For this analysis, muons are detected in three separate subdetectors.
Directly outside of the calorimeter, four-layer stacks of planar drift
chambers (CMU) detect muons with $p_T > 1.4$ GeV/c which penetrate the
five absorption lengths of the calorimeter~\cite{cmu}.  Farther out,
behind another 60~cm of steel, an additional four layers (CMP) detect
muons with $p_T > 2.0$ GeV/c~\cite{cmp}. The two systems cover the
same part of the central region $|\eta| \leq 0.6$, although the CMU
and CMP have different structures and their geometrical coverages do
not overlap exactly.  Muons between $ 0.6 \leq |\eta| \leq 1.0$ pass
through at least four drift layers lying on a conic section outside of
the central calorimeter; this system (CMX) completes the coverage over
the full fiducial region of the COT tracker~\cite{cmp}.  The presence
of a penetrating muon is reconstructed as a line segment or ``stub''
in one of the four-layer stacks.  Muon candidates are then identified
as isolated tracks which extrapolate to the stubs.  A track which is
linked to both CMU and CMP stubs is called a CMUP muon.
 
\subsection{\label{sec:beamposition}Beam Positions and the Primary
  Interaction Vertex} 

The event selection depends on reconstructing secondary vertices from
$b$ hadron decays. The identification of these decay vertices requires a
precise measurement of the primary vertex, the point from which all
prompt tracks originate. The primary vertex location in a given event
can be found by fitting well-measured tracks to a common point of
origin. 

The locus of all primary vertices defines the ``beamline,'' the
position of the luminous region of the beam-beam collisions through
the detector. The beamline can be used as a constraint to refine the
knowledge of the primary vertex in a given event. The first estimate
of the primary vertices ($x_V, y_V, z_V$) is binned in the $z$
coordinate.
A linear fit to $(x_V, y_V)$ {\it vs.} $z_V$ yields the beamline of
each run section.

The luminous region is long, with $\sigma_z = 29$ cm. The transverse
cross section is circular, with a width of approximately $30\,\mu$m at $z=0$,
rising to $\approx 50-60\,\mu$m at $|z|$ = 40 cm. The beam is neither
parallel to nor centered in the detector. At $z=0$, the beamline is at
$(x_V, y_V) \approx (-2.0, 3.9)$ mm, and has a slope of $\approx
5.0\,\mu$m/cm in the horizontal plane and $\approx 1.7\,\mu$m/cm in
the vertical plane. These parameters are rather stable, varying from 
their mean positions by  
no more than $\approx 20\%$ during periods of continuous data taking.

At high luminosities, more than one collision can occur on a given
bunch crossing; the primary vertices of the collision are typically
separated in the $z$ coordinate.  For the data analyzed here, there
are an average of 1.4 reconstructed vertices per event. The $z$ position
of each vertex is calculated from the weighted average of the $z$
coordinates of all tracks within 1 cm of a first iteration vertex,
with a typical resolution of $100\,\mu \rm{m}$.

A final determination uses all of the information above to recalculate
a best primary vertex in each candidate event for the $b$-tagging
procedure. This precise calculation, using a beam constraint and OI
tracks, is described fully in Sec.~\ref{sec:secvtx}. As part of the
lepton + jets event selection, the events are required to have the
reconstructed primary vertex located inside the luminous region
($|z|<60$ cm).

%% file: analysis_overview.tex
\section{\label{sec:data}Data Samples and Event Selection}

\subsection{Colliding Beam Data}

The colliding beam data used in this analysis were recorded during the
period March 2002 - August 2003, when the instantaneous Tevatron
luminosity ranged from 0.5 to 4.0~$\times 10^{31} ~{\rm cm^{-2}
  s^{-1}}$.  


Cherenkov light detectors in the very forward region ($|\eta|\geq
3.7$) record information on the instantaneous and total integrated
luminosity of the Tevatron~\cite{clc}.  The total integrated
luminosity for this period is $193 \pm 12~{\rm pb}^{-1}$; after quality
requirements on the silicon tracking, the data sample used for this
analysis amounts to $162 \pm 10~{\rm pb}^{-1}$ for CEM electrons and CMUP muons,
and $150 \pm 9~{\rm pb}^{-1}$ for CMX muons.

For the primary data samples used in this analysis, the detector is
triggered on high momentum electrons and muons. The electron hardware
triggers require an XFT track with $p_T \geq 8 ~{\rm GeV/c}$ matched
to an EM cluster with $E_T \geq 16$ GeV and the ratio of hadronic to
electromagnetic energy less than 0.125. The muon hardware triggers
require an XFT track with $p_T \geq 8 ~{\rm GeV/c}$ matched to muon
stubs in the joint CMUP configuration or in the CMX.  A complete
version of the offline lepton selection is performed online in the
last stage of triggering, and repeated in offline processing with
updated calibration constants.  Other secondary datasets described in
Sec.~\ref{sec:secvtx} use a jet trigger with a certain $E_T$ threshold
or an electron trigger with relaxed $E_T$ requirements.

\subsection{Monte Carlo Samples}

The understanding of acceptances, efficiencies, and backgrounds relies on 
detailed simulation of physics processes and the detector response. 
Most measurements of acceptance and efficiency rely on 
{\sc pythia}~v6.2~\cite{pythia} or {\sc herwig}~v6.4~\cite{herwig1,herwig2}.
These generators employ leading order matrix elements for the hard parton 
scattering, followed by parton showering to simulate gluon radiation and 
fragmentation. Each generator is used in conjunction with the CTEQ5L 
parton distribution functions~\cite{cteq}. For heavy flavor jets, we 
interface to {\sc qq}~v9.1~\cite{qq} to provide proper 
modeling of $b$ and $c$ hadron decays. 

The estimate of the $b$-tagging backgrounds due to higher order QCD
processes such as $Wb\bar b$ requires special care. 
This study of backgrounds in the $b$-tagged sample uses the
{\sc alpgen} program~\cite{alpgen}, which generates high multiplicity
partonic final states using exact leading-order matrix elements. The
parton level events are then passed to {\sc herwig} and {\sc qq} for
parton showering and $b$ and $c$ hadron decay.  Further discussion of
{\sc alpgen} can be found in Sec.~\ref{sec:hffracs}.

The CDF II detector simulation reproduces the response of the detector
to particles produced in $p\bar p$ collisions.  The same detector
geometry database is used in both the simulation and the
reconstruction, and tracking of particles through matter is performed
with {\sc geant3}~\cite{geant3}. Charge deposition in the silicon
detectors is calculated using a simple geometrical model based on the
path length of the ionizing particle and an unrestricted Landau
distribution.  The drift model for the COT uses a parameterization of
a {\sc garfield} simulation, with the parameters tuned to match COT
data~\cite{cot}.  The calorimeter simulation uses the {\sc
  gflash}~\cite{gflash} parameterization package interfaced with {\sc
  geant3}.  The {\sc gflash} parameters are tuned to test beam data
for electrons and high-$p_T$ pions, and they are checked by comparing
the calorimeter energy of isolated tracks in the collision data to
their momenta as measured in the COT.  Further detail on the CDF II
simulation can be found elsewhere~\cite{paulini}.
  
\subsection{$W$ + Jets Selection}

The selection identifies events consistent with the $W$ + jets
signature containing a high-momentum electron or muon (hereafter
referred to as ``lepton,'' $\ell$), large missing transverse energy, and
hadronic jets.  The event selection is summarized below.

The offline electron selection requires an EM cluster with $E_T \geq
20$ GeV matched to a track with $p_T \geq 10\,\mathrm{GeV}/c$.  The
cluster is required to have an electromagnetic fraction and shower shape
consistent with an electron deposit. The extrapolated track is
required to match the shower location as measured in the shower
maximum strip detector, and to have a momentum consistent with the
shower energy. Finally, since the electron from $W$ decay is expected
to be isolated from other energy deposits in the calorimeter, the
energy in a cone of radius $\Delta R = 0.4$ around the electron
cluster, but not including the cluster itself, is measured, and the
isolation ratio of the energy in the cone to the energy of the
electron is required to be less than 0.1.

Photon conversions in the detector material are a source of electron
backgrounds.  A conversion is defined as a pair of tracks (one of them
the electron) satisfying the following cuts:
\begin{itemize}
\item oppositely charged,
\item $|\Delta(xy)|<2\,\mathrm{mm}$, and 
\item $|\Delta(\cot\theta)|<0.04$,
\end{itemize}
\noindent where $\Delta(xy)$ is the distance between the tracks in the
$r-\phi$ plane at the point where they are parallel in that plane, and
$\Delta(\cot\theta)$ is the difference between the cotangents of the 
polar angles of the two tracks. 
Electrons that are part of an identified conversion 
pair are not considered further in the electron selection. 

The offline muon selection requires a COT track with $p_T \geq
20\,\mathrm{GeV}/c$ matched to a CMUP or CMX muon stub.  The matching
is based on the extrapolated track position at the chambers, accounting
for the effects of multiple scattering. The
energy in the calorimeter tower containing the muon is required to be
consistent with the deposition expected from a minimum ionizing
particle.  Backgrounds from cosmic rays are removed by requiring that
the track extrapolates to the origin, and that the minimum ionizing
tower energy deposit is within a narrow timing window around the beam
crossing.

In these high momentum lepton samples, the signal of the neutrino from
$W \rightarrow \ell\nu$ is large missing transverse energy, $\met$.
The $\met$ is calculated as the vector sum of the energy in
each calorimeter tower multiplied by the azimuthal direction of the
tower. If isolated high momentum muons are found in the event, the
$\met$ is corrected by subtracting the muon energy in the calorimeter
and adding the muon $p_T$ to the vector sum.  The selection finally
requires $\met \geq 20\,\mathrm{GeV}$.

In addition to the direct $t \rightarrow e \nu_e b$ and
$t \rightarrow \mu \nu_{\mu} b$ modes, this event selection has a 
small acceptance for top final states with $W \rightarrow \tau \nu$ and
a subsequent leptonic $\tau$ decay, or with high momentum semi-leptonic 
b quark decays. 
These are included in the signal acceptances calculated in Sec. IX. 

$Z$ bosons and top dilepton decays that contribute to the inclusive
high $p_T$ lepton dataset are removed by flagging the presence of a
second lepton.  Any event with two leptons satisfying the lepton
identification is removed, as well as those events where the second
lepton is an electron in the plug calorimeter or a muon that fails the
CMUP requirement, but has one CMU or CMP muon segment.  Finally, we attempt
to remove $Z$ bosons without a well identified second lepton by eliminating 
events with one lepton and certain second objects which form an invariant 
mass between 76 and 106~GeV/$c^2$ with the primary lepton. For primary 
muons the other object is an
opposite-signed isolated track with $p_T > 10$ GeV/c. For primary
electrons the second object may be such a track, an electromagnetic
cluster, or a jet with $E_T > 15$ GeV and $|\eta| \leq 2.0$ that has
fewer than three tracks and an electromagnetic energy fraction greater
than 95\%.  The correction for the residual $Z$ boson contribution to
the $W$ + jets sample is described in Section~\ref{sec:checks}. Small
contributions from $Z \rightarrow \tau \bar{\tau}$ where a $\tau$ is tagged 
are treated as a separate background, and described in Section~\ref{sec:m2bkg}.

\begin{figure}
  \includegraphics[width=0.5\textwidth]{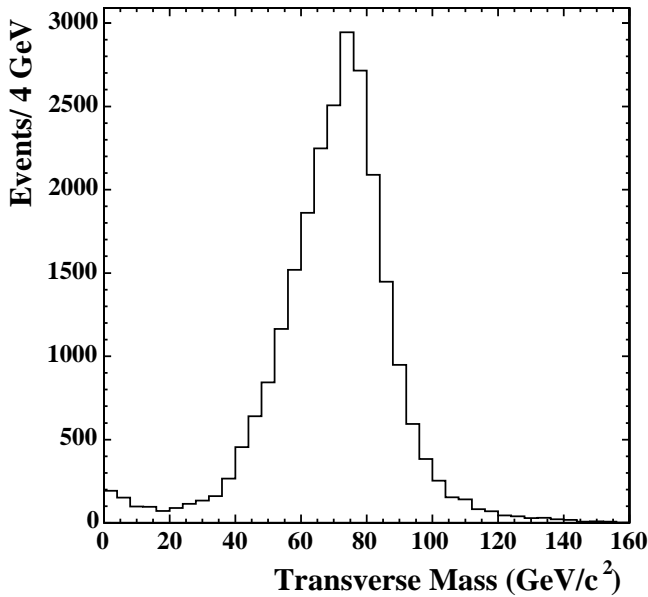}%
  \caption{\label{fig:wmass}Transverse mass of the identified lepton
    and inferred neutrino, consistent with $W$ boson production 
(162 $\mathrm{pb}^{-1}$ data sample).}
\end{figure}

The number of jets produced in association with the leptonically
decaying $W$ in the event is measured by selecting jets of cone radius
$\Delta R = 0.4$, with $E_T \geq 15\,\mathrm{GeV}$ and $|\eta| \leq
2.0$.  The jets are clustered after removing towers associated with
the selected isolated electron from the leptonic $W$ decay, and after
correcting the tower $E_T$ for the location of the primary vertex $z$
coordinate. The number of events in each jet multiplicity bin is shown
in Table~\ref{tab:pretag}.  The overall ${\rm acceptance \times
  efficiency}$ of this selection for $t\bar t$ events in the
lepton+jets channel with three or more jets, including the leptonic
branching ratios, is roughly 4\% for the electron channel, 2\% for
muons in the CMUP, and 1\% for muons in the CMX.

The presence of the $W$ boson in the selected events is verified by
calculating the transverse mass of the lepton and the missing energy:
$M_T = \sqrt{(E_T(\ell)+E_T(\nu))^2 -
(\overrightarrow{P_T}(\ell)+\overrightarrow{P_T}(\nu))^2}$. The
distribution of this variable for all events passing the requirement
  of a lepton, missing energy, and at least one jet is shown in
  Fig.~\ref{fig:wmass}, and displays the Jacobian edge associated with
  $W$ production and decay.

\begin{table*}
  \caption{\label{tab:pretag}Number of events selected, before
    $b$-tagging, for each jet multiplicity.}
  \begin{ruledtabular}
    \begin{tabular}{ccccccc}
      & $W$ + 1 jet & $W$ + 2 jets & $W$ + 3 jets & $W$ + 4 jets & $W$ + 3 jets & $W$ + 4 jets \\ 
      & & & \multicolumn{2}{c}{$H_T > 0$} & \multicolumn{2}{c}{$H_T >
        200$~GeV} \\ \hline
      Electrons & 8828 & 1446 & 241 & 70 & 117 & 63 \\
      Muons & 6486 & 1002 & 146 & 37 & 63 & 28 \\
    \end{tabular}
  \end{ruledtabular}
\end{table*}

As a final optimization step, the selection will incorporate an
additional cut on the total transverse energy $H_T$ of all objects in
the event. Events from $t\bar t$ production have, on average, a
significantly greater total transverse energy than background events.
The optimization of this requirement and acceptance
corrections and uncertainties will be discussed in
Sec.~\ref{sec:optimization}.

Because the $t\bar t$ signal is expected to contribute significantly to
the sample of events with $W$ + 3 jets or $W$ + $\geq 4$ jets, an excess
of observed events over the expected background with those jet multiplicities 
is assumed to be entirely due to $t\bar t$ production.  The observed results
for events with $W$ + 1 jet or $W$ + 2 jets, where the $t\bar t$ contribution
is negligible, serve as a check of the background prediction. 

In Reference~\cite{chicago}, the momentum spectrum of the leading jets in 
the  $W$ + 1 and 2 jet events is shown to be a reasonable model of the
backgrounds in the  $W$ + 3 or 4 jet events, and is used in deriving a 
completely independent estimate of the b-tag backgrounds to top
production in the $W$ + 3 or 4 jet channels. The estimated background, 
$18 \pm 4$ events, is in good agreement with our overall  
estimate of $23 \pm 3$ (for $H_T>0$), derived from an explicit calculation 
for each contributing background process (see Sec.\ref{sec:m2bkg}). The 
independence of these background estimates allows for a combined 
cross section calculation which will appear in a future paper. 
We note here that the technique of Reference~\cite{chicago} will
work only in the study of top quarks in the $W$ plus jet mode, and we 
consider it a cross-check on our more general technique for 
calculating b-tag backgrounds, which will be employed in other 
b-tagging analyses at CDF.

The final cross section calculation, $\sigma_{t\bar t} = (N_\mathrm{obs} -
N_\mathrm{bkg})/(\epsilon_{t\bar t} \times \mathcal{L})$, depends on the
product $\epsilon_{t\bar t}$ of signal acceptance and selection efficiency, the
expected number of non-$t\bar t$ background events $N_\mathrm{bkg}$, and
the integrated luminosity $\mathcal L$.

%% file: secvtx_studies.tex
\section{\label{sec:secvtx}Secondary Vertex $b$-Tagging}


In this Section we describe and discuss the performance of an
algorithm to identify jets resulting from heavy quark ($b$,$c$)
fragmentation.  This ``SecVtx'' algorithm is based upon the algorithm
used to discover the top quark~\cite{run1prd}.  Most of the non-$t\bar
t$ processes found in the $W$ + jets sample do not contain heavy
quarks in the final state.  Requiring that one or more of the jets in
the event be tagged by SecVtx keeps more than half of the $t\bar t$
events while removing approximately 95\% of the background.

\subsection{\label{sec:algorithm} Description of the SecVtx Algorithm }

The SecVtx algorithm relies on the displacement of secondary vertices
relative to the primary event vertex to identify $b$ hadron decays.
The Run~II algorithm is essentially unchanged from
Run~I~\cite{run1prd}, but the track selection cuts have been re-tuned
for the CDF~II detector.



In order to select displaced tracks coming from decays of long-lived
hadrons, precise knowledge of the collision point is necessary.  To
find an event-by-event primary vertex, we first identify which of the
vertices described in Section~\ref{sec:detector} is nearest the
identified high-momentum electron or muon.  For other datasets without
high-momentum leptons, we use the vertex which has the highest total
scalar sum of transverse momentum of associated tracks.  The position
of the primary vertex is then determined by fitting together the
tracks within a $\pm 1\,\mathrm{cm}$ window in $z$ around this vertex.
The procedure starts by fitting a vertex using all tracks within the
$z$ window and with impact parameter significance (relative to the
beamline) $|d_0/\sigma_{d_0}|<3$, where $\sigma_{d_0}$ includes the
uncertainty on both the track and the beamline positions.  The
transverse profile of the beamline at the $z$ of the original vertex
estimate is also used as a constraint in the fit.  A pruning stage
removes tracks which contribute $\chi^2 > 10$ to the fit (or the track
with the largest $\chi^2$ contribution if the total fit reduced
chi-squared per degree of freedom $\chi^2/\mathrm{ndf} > 5$).  After
the initial pruning, the fit is repeated using only the remaining
tracks until a vertex with no tracks over the $\chi^2$ cut is found.
If no tracks survive the pruning stage then the beamline profile is
used for the primary vertex position estimate.  In the event sample
used for these results the uncertainty in the fitted transverse
position ranges from $10-32\,\mu$m depending upon the number of
reconstructed tracks and the topology of the event.

Secondary vertex tagging operates on a per-jet basis, where only
tracks within the jet cone are considered for each jet in the event.
A set of cuts involving the transverse momentum, 
the number of silicon hits attached to the
tracks, the quality of those hits, and the $\chi^2/\mathrm{ndf}$ of
the final track fit are applied to reject poorly reconstructed tracks.
Only jets with at least two of these good tracks can produce
a displaced vertex; a jet is defined as ``taggable'' if it has two
good tracks.  Displaced tracks in the jet are selected based on the
significance of their impact parameter with respect to the primary
vertex and are used as input to the SecVtx algorithm.  SecVtx uses a
two-pass approach to find secondary vertices.  In the first pass,
using tracks with $p_T > 0.5$ GeV/$c$ and $|d_0/\sigma_{d_0}|>2.5$,
it attempts to reconstruct a secondary vertex which includes at least
three tracks (at least one of the tracks must have
$p_T>1\,\mathrm{GeV}/c$).  If the first pass is unsuccessful, it
performs a second pass which makes tighter track requirements ($p_T >
1\,\mathrm{GeV}/c$ and $|d_0/\sigma_{d_0}|>3$) and attempts to
reconstruct a two-track vertex (one track must have
$p_T>1.5\,\mathrm{GeV}/c$).

Once a secondary vertex is found in a jet, the two-dimensional decay
length of the secondary vertex $L_{2D}$ is calculated as the
projection onto the jet axis, in the $r-\phi$ view only, of the
vector pointing from the primary vertex to the
secondary vertex.  The sign of $L_{2D}$ is defined relative to the jet
direction, specifically by the absolute difference $|\phi|$ between
the jet axis and the secondary vertex vector (positive for $<
90^\circ$, negative for $> 90^\circ$).  Secondary vertices
corresponding to the decay of $b$ and $c$ hadrons are expected to have
large positive $L_{2D}$ while the secondary vertices from random
mis-measured tracks are expected to be less displaced from the primary
vertex.  To reduce the background from the false secondary vertices
(mistags), a good secondary vertex is required to have
$L_{2D}/\sigma_{L_{2D}} > 3$ (positive tag) or $L_{2D}/\sigma_{L_{2D}}
< -3$ (negative tag), where $\sigma_{L_{2D}}$, the total estimated
uncertainty on $L_{2D}$ including the error on the primary vertex, is
estimated vertex-by-vertex but is typically $190\,\mu$m.  The negative
tags are useful for calculating the false positive tag rate, as
detailed in Section~\ref{sec:mistag}.  A tagged jet is defined to be a
jet containing a good secondary vertex (the SecVtx algorithm will find
at most one good vertex per jet).

\subsection{\label{sec:efficiency}Measurement of Tagging Efficiency} 

The results described in this paper require a knowledge of the tagging
efficiency for $t\bar t$ events, \textit{i.e.}, how often at least one
of the jets in a $t\bar t$ event is positively tagged by SecVtx.
Because it is not possible to measure this directly in $t\bar t$
events we have adopted a different strategy.  A sample of jets whose
heavy flavor fraction can be measured is used to derive the per-jet
tagging efficiency in the data for that sample.  The heavy flavor in
this sample is a mixture of charm and bottom, with the relative
proportions of each determined from the mass spectrum of SecVtx tagged
jets and the ratio of charm/bottom tagging efficiencies predicted by
the Monte Carlo simulation.  The charm component is suppressed by
requiring a second tagged jet in the event, so that the measured tag
efficiency is dominated by the contribution from bottom.  Because the
jets in $t\bar t$ events will in general have different energies,
pseudorapidities, and track multiplicity than the jets in the
calibration sample, the measured efficiency cannot be used directly.
Instead, a matching sample of Monte Carlo jets is used to determine
the tagging efficiency in the simulation for jets like those in the
calibration sample, and the ratio of efficiencies between data and
simulation (scale factor) is then used to correct the tagging
efficiency in $t\bar t$ Monte Carlo samples.  In other words, the
geometrical acceptance and energy dependence of the tagger are taken
from the simulation, with the overall normalization determined from
the data.

To measure the efficiency for tagging heavy flavor hadrons, we use a
sample of low-$p_T$ inclusive electron data which is enriched in
semileptonic decays of bottom and charm hadrons.  For the matching
Monte Carlo sample we use the {\sc herwig}~\cite{herwig1} program to
generate 2$\rightarrow$2 parton events, which are passed through a
filter requiring an electron with $p_T > 7\,\mathrm{GeV}/c$ and
$|\eta| < 1.3$.  Events passing this filter are processed using the
detector simulation described in Section~\ref{sec:detector}.


Electrons in the events are identified using the selection in
Section~\ref{sec:data}, except with lower thresholds $E_T>9\,\mathrm{GeV}$
and track $p_T>8\,\mathrm{GeV}/c$.  Further differences from
Section~\ref{sec:data} are that the electrons are required to be
non-isolated and conversions are not removed.  The electron track must
also pass through every layer of the SVX detector.

\begin{figure}[htbp]
  \begin{center}
    \includegraphics*[width=0.5\textwidth]{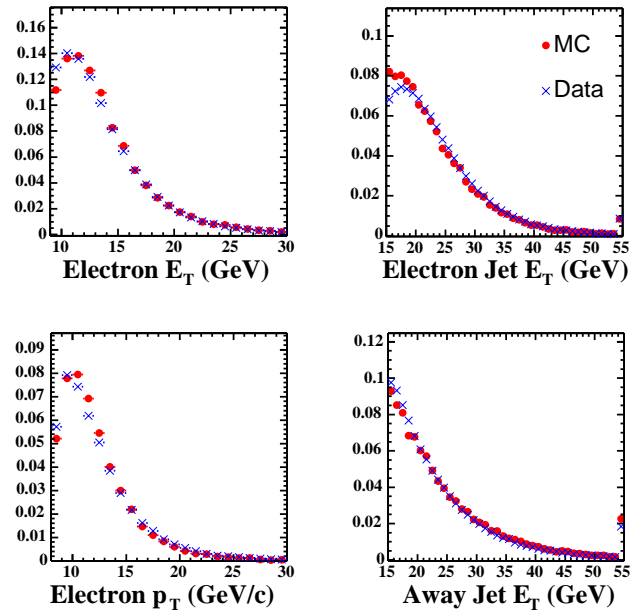}
    \caption{Data/Monte Carlo comparison of some quantities of tagged
      electron jets ($L_{2D}>0$, identified conversions have been
      removed for plotting purposes). Histograms are normalized to unit area.
From top-left, clockwise: electron $E_T$, electron-jet $E_T$,
      away-jet $E_T$, electron $p_T$.  (The last bin includes all
      overflow entries.)
    \label{fig:etcomp}}
  \end{center}
\end{figure}

\begin{figure}[htbp]
  \begin{center}
    \includegraphics*[width=0.5\textwidth]{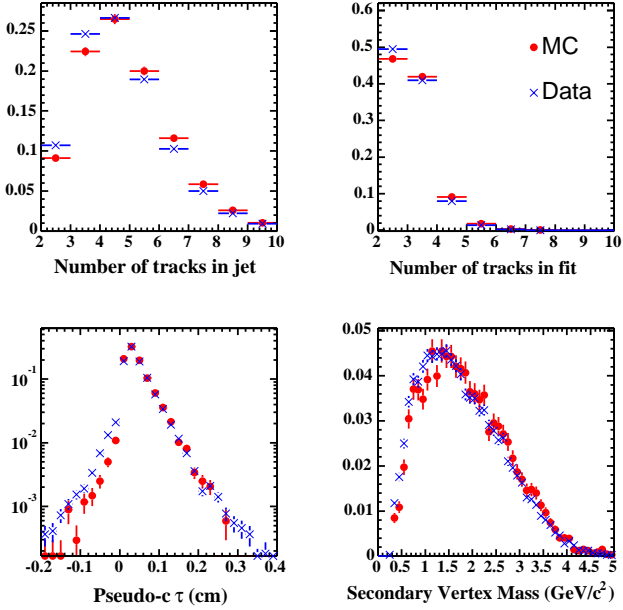}
    \caption{Data/Monte Carlo comparison of some quantities of
      tagged electron jets (identified conversions have been removed
      for plotting purposes). Histograms are normalized to unit area.
      From top-left, clockwise: number of good tracks in the jet,
      number of tracks in the tagged
      vertex, vertex mass of positively tagged electron-jets;
     pseudo-$c\tau$ of (positively or negatively) tagged
      electron-jets.
    \label{fig:secvtxcomp}}
  \end{center}
\end{figure}

Along with the electron we require two jets, the ``electron jet'' and
the ``away jet.''  The electron jet is required to have $E_T > 15$ GeV
(including the energy of the electron) and to be within 0.4 of the
electron in $\eta-\phi$ space (in other words the electron is within
the jet cone), and is presumed to contain the decay products of a
heavy flavor hadron.  The away jet is required to have $E_T > 15$ GeV
and $|\eta| < 1.5$, and it must be approximately back-to-back with the
electron jet ($\Delta \phi > 2\,\mathrm{rad}$).  A total of 481,301
events of the data sample pass these event selection requirements.
Figures~\ref{fig:etcomp} and \ref{fig:secvtxcomp} show that the Monte
Carlo is an adequate representation of the data sample for relevant
event selection and tagging variables.  The differences can be
attributed to the presence of fake electrons in the data which are not
completely removed even after requiring a SecVtx tag, and which are
not present in the Monte Carlo due to the generator-level electron
filter.  The discrepancy in Figure~\ref{fig:secvtxcomp} on the
negative side of the pseudo-$c\tau$ plot shows that the Monte Carlo
underestimates the mistag rate observed in the data.


In order to measure the tagging efficiency for electron jets, it is first
necessary to characterize their heavy flavor content.  Two methods
are used to measure the fraction $F_b$ of electron jets
which contain a $b$ hadron.  The first method is to reconstruct $D^0
\rightarrow K^-\pi^+$ decays within the electron jet and use the
invariant mass sidebands to subtract background; this method yields $F_b =
0.139 \pm 0.021$.  The second method involves searching for secondary
muons within the electron jet resulting from cascade $c$ decays using
the same-sign rate to estimate the background; this method gives $F_b = 0.228
\pm 0.037$.  Because the agreement is only at the 2$\sigma$ level, the
uncertainty on the weighted average is inflated by 2.09 based on the $\chi^2$
of the two determinations.  The combined result of the two
measurements is $F_b = 0.161 \pm 0.038$.

\begin{figure}[htbp]
  \begin{center}
    \includegraphics*[width=0.5\textwidth]{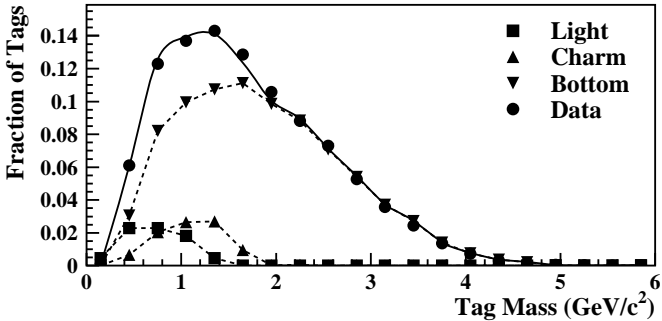}
    \caption{Fit (solid line) of the relative $b$ and $c$ contributions to the vertex
      tag mass distribution.  Templates for the different flavors are
      derived from simulation (and the data in the case of light flavor).
      The error bars for the data are contained within the markers.}
    \label{fig:massfit}
  \end{center}
\end{figure}

The fraction $F_c$ of electron jets which came from a charm quark also
contributes to the total heavy flavor fraction $F_\mathrm{HF} = F_b +
F_c$.  An estimate of the amount of $c$ relative to $b$ in the
electron jet is derived from a fit to the invariant mass spectrum of
the tracks in the positive tags found in the electron jets.  Templates
for $b$, $c$, and light-flavor jets taken from the Monte Carlo (and
also from the data for light-flavor) were fitted to the distribution,
as shown in Figure~\ref{fig:massfit}, to obtain the ratio of $c$ to
$b$ after requiring a positive tag.  The result of this fit is
$F_c^\mathrm{tag}/F_b^\mathrm{tag} = 0.118 \pm 0.017$, where the
uncertainty is dominated by the systematic error due to varying the
light-flavor template.  A value for $F_c/F_b$ before any tagging is
obtained by multiplying this result by the ratio of tagging
efficiencies $\epsilon_b/\epsilon_c = 5.2 \pm 0.4$ predicted by the
Monte Carlo, resulting in $F_c/F_b = 0.61 \pm 0.10$.  The uncertainty
on $\epsilon_b/\epsilon_c$ is derived from reweighting the Monte Carlo
to match the multiplicity of tracks in the jet passing the quality cuts
which is observed in the data.  Applying the factor of 0.61, the total
heavy flavor fraction of electron jets $F_\mathrm{HF}$ is $0.259 \pm
0.064$.

To measure the tagging efficiency of the heavy flavor electron jets we
employ a double-tag technique, requiring that the away jet be
tagged by SecVtx.  This enhances the heavy flavor fraction of the
electron jets and reduces the dependence on $F_\mathrm{HF}$, which we
were only able to constrain at the 25\% level.  Another benefit of the
double-tag is to reduce the influence of the charm component, so that
the resulting heavy flavor tag efficiency is more representative of
the $b$-tagging efficiency.  Tagging the away jet reduces the charm from 61\% of the bottom component down to around 10\%.  The
difference in the tag efficiency for semileptonic decays, which we
measure, and generic heavy hadron decays is used later to estimate a
systematic error.

The tagging efficiency for heavy flavor jets containing an electron, derived
from the numbers of double- and single-tags, is
\begin{equation}
  \label{eq:eff_doubletag}
  \varepsilon =
  \frac{ (N^\mathrm{e+}_\mathrm{a+}-N^\mathrm{e-}_\mathrm{a+})-
    (N^\mathrm{e+}_\mathrm{a-}-N^\mathrm{e-}_\mathrm{a-})}
  {(N_\mathrm{a+}-N_\mathrm{a-})}\cdot \frac{1}{F_\mathrm{HF}^\mathrm{a}},  
\end{equation}
\noindent where $N_\mathrm{a+}$ and $N_\mathrm{a-}$ are the numbers of positive
and negative tagged away jets, and $N^\mathrm{e+}_\mathrm{a+}$, for example,
is the number of events where both electron and away jet are positive
tagged.

The factor $F_\mathrm{HF}^\mathrm{a}$ is the fraction of electron jets
containing heavy flavor for events where the away jet is tagged.  This
number is less than one due to events where the away jet is
mistagged or contains heavy flavor due to gluon splitting or flavor
excitation, and the electron is either a fake or part of a photon
conversion pair.  In order to estimate these effects we use
identified conversions (see Section~\ref{sec:data}) to probe the light
flavor composition of the electron jets.  In this way we write
$F_\mathrm{HF}^\mathrm{a}$ as
\begin{equation}
  \label{eq:bfrac_HFeffect}
  F_\mathrm{HF}^\mathrm{a} = 1-
  \frac{\left(\frac{N_\mathrm{c}^\mathrm{a+}-N_\mathrm{c}^\mathrm{a-}}
      {N_\mathrm{a+}-N_\mathrm{a-}}-\epsilon_{c}'\right)}
  {\frac{N_\mathrm{c}}{N}-\epsilon_{c}'}
  \cdot (1-F_\mathrm{HF}),  
\end{equation}
\noindent where $N$ is the number of events passing the selection, 
$\epsilon_{c}' = \frac{N_\mathrm{c}^\mathrm{e+}-N_\mathrm{c}^\mathrm{e-}}
        {N_\mathrm{e+}-N_\mathrm{e-}}$, and the c subscript refers
to events where the electron was identified as a conversion.  A full 
derivation of this expression can be found in the Appendix~\ref{app:HFeffect}.

\begin{figure}[htbp]
  \begin{center}
    \includegraphics*[width=0.5\textwidth]{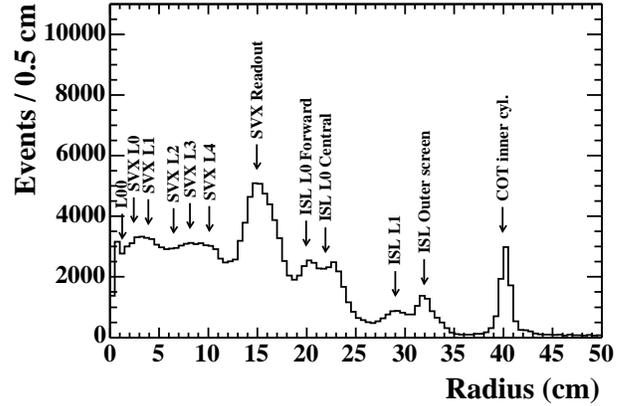}
    \caption{Radius of identified conversions in data, with location
      of the silicon detector layers (L00, SVX and ISL), readout system,
      and ISL and COT main mechanical structures.}
    \label{fig:conversions}
  \end{center}
\end{figure}

To illustrate the effectiveness of the conversion finder,
Figure~\ref{fig:conversions} shows the estimated radius of the conversion
point for identified pairs.  Peaks corresponding to known detector
structures are clearly visible. 

\begin{figure}[htbp]
  \includegraphics*[width=0.5\textwidth]{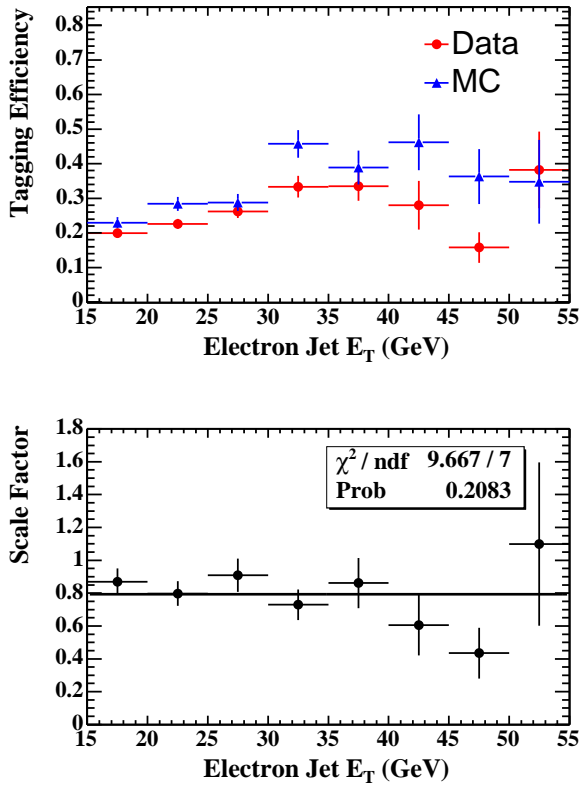}
  \caption{Efficiency to tag a heavy flavor electron jet as a function
  of jet $E_T$ in data and Monte Carlo (top), and data/MC scale factor
  (bottom). The scale factor is consistent with being constant over
  the $E_T$ range investigated.  The solid line shows the result of
  the fit to the binned scale factor from which the $\chi^2$ was
  derived.  It is consistent with although not identical to the value
  of 0.82 obtained for the overall sample.}

  \label{fig:eff_doubletag}
\end{figure}

We use the $F_\mathrm{HF}$ value derived above for data, a value of
$F_{\mathrm{HF}}^{\mathrm{MC}} = 0.861$ for the Monte Carlo (found by
counting jets which are matched to a heavy quark), and
Equations~\ref{eq:eff_doubletag}~and~\ref{eq:bfrac_HFeffect} (see
Appendix) to calculate the efficiencies to tag a heavy flavor jet
containing an electron in data and Monte Carlo.  The resulting values
averaged over jet $E_T$ are given in Table~\ref{tab:eff_doubletag}.
The efficiencies as a function of the $E_T$ of the jet are shown in
Figure~\ref{fig:eff_doubletag}.  The ratio of data to Monte Carlo
efficiencies (scale factor) is also shown as a function of $E_T$.
Additionally, a sample of jet data with one jet having $E_T > 50$ GeV
and a corresponding 2$\rightarrow$2 {\sc Pythia} Monte Carlo sample have been
used to determine that the ratio of jet tag rates is flat over a wider
jet $E_T$ range than that spanned by the electron calibration sample.
These samples are also used to estimate a systematic uncertainty for
extrapolating the scale factor to the higher-$E_T$ jets (typically
40-120 GeV) characteristic of top quark decays.


\begin{table}[htbp]
    \caption{Efficiency to tag a heavy flavor electron jet in data and
      Monte Carlo, and the data/MC ratio (scale factor).
      Uncertainties on the efficiencies are statistical only;
      systematic uncertainties on the scale factor are summarized in
      Table~\ref{tab:sfsystematics}.} 
    \label{tab:eff_doubletag}
  \begin{ruledtabular}
    \begin{tabular}{cc}
      $\varepsilon (\mathrm{Data})$ & $0.240 \pm 0.007$ \\
      $\varepsilon (\mathrm{MC})$ & $0.292 \pm 0.010$ \\
      Scale Factor                & $0.82 \pm 0.06$
    \end{tabular}
  \end{ruledtabular}
\end{table}


Several sources of systematic uncertainty have been considered and are
summarized in Table~\ref{tab:sfsystematics}.  The
$F_\mathrm{HF}^\mathrm{a}$ method uncertainty accounts for assumptions
made in the calculation of $F_\mathrm{HF}^\mathrm{a}$ about the
tagging efficiency of heavy flavor electron jets containing a
conversion electron pair.  The mistag subtraction uncertainty is
related to the asymmetry in negative tags {\em{vs.}} fake positive
tags described in the next subsection, and is conservatively estimated
by scaling the negative tag rates for all jets by either zero (no
subtraction) or by a factor of two.  $E_T$ dependence was described
earlier, and the $B$-decay uncertainty allows for a possible
difference in the scale factor due to the lower charged particle
multiplicity of semileptonic $B$ decays compared to all possible decay
modes.  Combining all systematic and statistical errors we obtain a
data to Monte Carlo tagging efficiency scale factor of $0.82 \pm
0.06$.

\begin{table}[htbp]
    \caption{Relative uncertainties on the data to Monte Carlo tagging
efficiency scale factor, in percent.}
    \label{tab:sfsystematics}
  \begin{ruledtabular}
    \begin{tabular}{lc}
      Source & uncertainty (\%) \\
      \hline
      $F_\mathrm{HF}$ & 3.5 \\
      $F_\mathrm{HF}^\mathrm{a}$ method & 3.0 \\
      mistag subtraction & 3.0 \\
      $E_T$ dependence & 2.5 \\
      B-decay & 1.2 \\ \hline
      total systematic error & 6.2 \\ \hline
      data statistics & 3.2 \\
      MC statistics & 3.6 \\ \hline
      total uncertainty & 7.8
    \end{tabular}
  \end{ruledtabular}
\end{table}

A variation of the double-tag technique has also been studied which
uses the single-tag rate of electron jets rather than the measurements
of $F_\mathrm{HF}$.  First we write the efficiencies in the data as $\epsilon =
SF\times\epsilon_\mathrm{MC}$ and $\epsilon^\mathrm{single} =
SF\times\epsilon_\mathrm{MC}^\mathrm{single}$, where $\epsilon$ is defined in
Equation~\ref{eq:eff_doubletag} and $\epsilon^\mathrm{single} =
(N^{e+}-N^{e-})/(F_\mathrm{HF} N)$ is the net single-tag efficiency for
heavy-flavor electron jets.  Although both $\epsilon$ and
$\epsilon^\mathrm{single}$ are tag efficiencies for heavy-flavor jets, they
generally differ because the requirement of an away jet tag suppresses
the charm content of the sample relative to bottom.

Substituting for $\epsilon$ and $F_\mathrm{HF}$ (using the relation
between $F_\mathrm{HF}$ and
$SF\times\epsilon_\mathrm{MC}^\mathrm{single}$) into
Equations~\ref{eq:eff_doubletag_full}~and~\ref{eq:bfrac_dtag} allows
solution for the efficiency scale factor $SF$ directly in terms of the
data tag and conversion rates, and of the MC tag efficiencies
$\epsilon_\mathrm{MC}$ and $\epsilon_\mathrm{MC}^\mathrm{single}$.  A
result of $SF = 0.81$ is obtained, consistent with the method
described above and with similar systematic and statistical errors.



\subsection{\label{sec:mistags}Measurement of the Mistag Rate} 

A ``mistag'' is defined to be a jet which did not result from the
fragmentation of a heavy quark, yet has a SecVtx secondary vertex.
Mistags are caused mostly by random overlap of tracks which are
displaced from the primary vertex due to tracking errors, although
there are contributions from $K_S$ and $\Lambda$ decays and nuclear
interactions with the detector material (the beampipe or the inner
silicon layers) as well.  Contributions from these effects are
measured directly from jet data samples without relying on the
detector simulation.

Because the SecVtx algorithm is symmetric in its treatment of $d_0$
and $L_{2D}$ significance, the tracking-related mistags should occur
at the same rate for $L_{2D}>0$ and $L_{2D}<0$.  Therefore, a good
estimate of the positive mistag rate due to resolution effects can be
obtained from the negative tag rate.  However, some of the negative
tags occur in jets which do contain heavy flavor, so that part must be
subtracted.  In addition, the negative rate will not reflect the
mistags due to lifetime or interactions with the detector material.
Corrections for all of these effects are determined using fits to the
pseudo-$c\tau$ spectra of tagged vertices, described in
Section~\ref{sec:hfdijetdata}.  The sum of these corrections is found to
be $20 \pm 10$\% of the negative tag rate, consisting of a subtraction
of 20\% for removal of the heavy flavor negative tags, and an addition
of 40\% to account for the mistags due to lifetime and material
interactions.





The rate of negative tags for taggable jets is measured in
an inclusive sample of jet triggers.  The rate is parameterized as a
function of four jet variables -- $E_T$, track multiplicity,
$\eta$, and $\phi$ -- and one event variable $\Sigma E_T$, the scalar
summed $E_T$ of all jets in the event with $E_T>10$ GeV and
$|\eta|<2.4$.  These parameterized rates are used to obtain the
probability that a given jet will be negatively tagged.

\begin{figure}[htbp]
  \begin{center}
    \includegraphics*[width=0.5\textwidth]{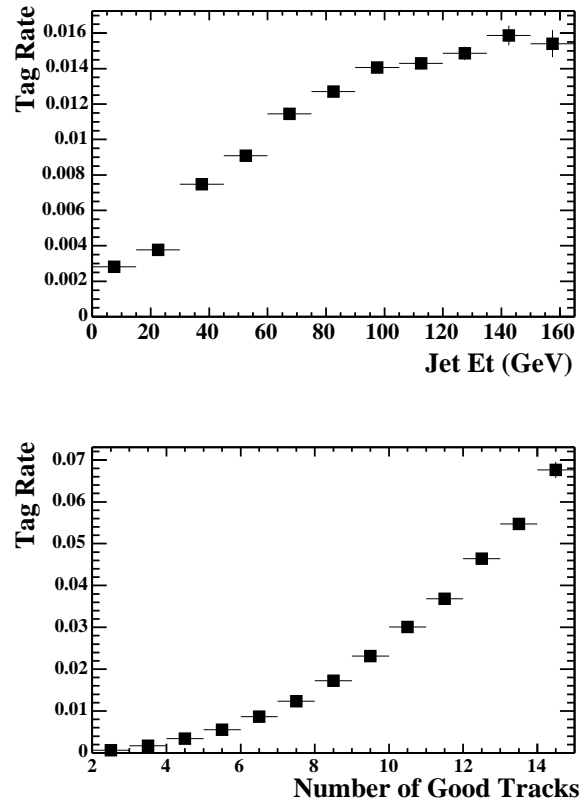}
    \caption{SecVtx negative tag rate as a function of jet $E_T$ and track
      multiplicity in the inclusive jet data.} 
    \label{fig:tagrate}
  \end{center}
\end{figure}

The full five-dimensional tag rate matrix was determined using
inclusive 20 GeV, 50 GeV, 70 GeV, and 100 GeV jet trigger samples, for a
total of $11.5 \times 10^6$ events.  Figure~\ref{fig:tagrate} shows the
negative tag rate per taggable jet as a function of jet $E_T$ and
track multiplicity (and integrated over the other variables) for all
the events in the inclusive jet sample.  These rates have not been
scaled by the $1.2 \pm 0.1$ correction discussed above which is
applied to convert to an estimate of the positive mistag rate.


Detailed cross checks were performed on the tag rate matrix to verify
its self-consistency and to check predictability and sample
dependence.  Both the total tag rates and the tag rates as functions
of various quantities were used to check how well the matrix predicts
the observed data and to estimate systematic errors.
Table~\ref{tab:mistag_checks} summarizes the differences between the
matrix predictions and the observed tag rates in various validation
samples.  The four jet trigger samples described above were used,
along with an independently-triggered sample requiring four jets with
$E_T>15\,\mathrm{GeV}$ and $\Sigma E_T > 125\,\mathrm{GeV}$, referred
to as the ``SumEt'' sample.  The table is divided into two sections.
Each row in the table compares the tag rate predicted from one sample
with the observed rate in a second, different sample.


\begin{table*}[htbp]
    \caption{Differences in predicted and observed negative tagging
      rates for various samples.  The first four rows with labels of
      the form Sample1-Sample2 compare observed tag rates in Sample~2
      to the rates predicted by a matrix made from Sample~1.  The last
      three rows compare the observed tag rates for trigger jets,
      non-trigger jets, and jets in the SumEt sample with predictions
      from the standard mistag matrix derived from all four jet
      samples.\label{tab:mistag_checks}}
  \begin{ruledtabular}
    \begin{tabular}{cccc}
             &  Observed Negative Tag Rate (\%) & Predicted Negative Tag Rate
             (\%) & Obs./Pred. \\ \hline  
      Jet20-Jet50  & 0.728$\pm$ 0.008 & 0.677$\pm$0.046 & 1.08$\pm$ 0.08 \\
      Jet50-Jet70  & 0.958$\pm$ 0.009 & 0.930$\pm$0.013 & 1.03$\pm$ 0.02 \\ 
      Jet50-Jet100 & 1.219$\pm$ 0.009 & 1.151$\pm$0.044 & 1.06$\pm$ 0.04 \\ 
      Jet50-SumEt  & 0.730$\pm$ 0.005 & 0.712$\pm$0.015 & 1.03$\pm$ 0.02 \\ \hline
      Trigger Jet  & 0.565$\pm$ 0.005 & 0.587$\pm$0.005 & 0.96$\pm$ 0.01 \\ 
      Non-Trigger Jet  & 0.659$\pm$ 0.005 & 0.640$\pm$0.006 & 1.03$\pm$ 0.01 \\
      SumEt     &  0.712$\pm$ 0.006 & 0.726$\pm$0.007 & 0.98$\pm$ 0.01 \\
    \end{tabular}
  \end{ruledtabular}
\end{table*}

The differences in the tag rates of trigger jets and non-trigger jets
are well predicted by the matrix. This is mostly due to the inclusion
of the jet $E_T$, $\eta$, and $\phi$ into the matrix binning. The
remaining residual difference is taken as a systematic error in the
final result. 


 
The systematic uncertainties assigned to the tag rate matrix
predictions are summarized in Table~\ref{tab:mistag_sys}. We assume
that the various contributions are uncorrelated and add them in
quadrature to find a total systematic uncertainty of 8\% on the
negative tag rates, which combined with the uncertainty on the
correction factor $1.2\pm0.1$ yields a total mistag rate relative
uncertainty of 11\%.

\begin{table}[htbp]
    \caption{Systematic uncertainties assigned to the negative tag
    rate matrix predictions.}
    \label{tab:mistag_sys}
  \begin{ruledtabular}
    \begin{tabular}{ccc}
      Source & Uncertainty \\ \hline 
      Trigger jet bias  & 4\% \\ 
      Sample bias       & 7\% \\ 
      Statistics        & 1\% \\ \hline  
      Total             & 8\% \\
    \end{tabular}
  \end{ruledtabular}
\end{table}

%% file: hf_fractions.tex
\section{\label{sec:hffracs}Heavy Flavor Contributions to $W$+Jets}

Heavy flavor production in association with a vector boson
({\em{e.g.}}  $Wb\bar b$, $Wc\bar c$, $Wc$) contributes significantly
to the non-$t\bar t$ background in the $b$-tagged lepton + jets
sample, even though $W$ + light flavor jet production dominates the
pretag sample.  Several Monte Carlo generators are capable of
performing matrix element calculations for $W/Z$ + jets, even to high
jet multiplicity, but these generators use leading-order calculations.
As a result, the overall normalization of these calculations has a
large theoretical uncertainty, even though the relative contributions
of the important diagrams are well-defined.

For this reason, the relative fraction of $W$ + heavy flavor
production is calculated in a matrix element Monte Carlo program, and
the overall normalization of the $W$ + jets production is measured
with collider data. The two results can be combined to estimate the
$W$ + heavy flavor background.

For this analysis, we use a new event generator, {\sc alpgen}~\cite{alpgen},
which calculates exact matrix elements at leading order for a large
set of parton level processes in QCD and electroweak interactions.
All heavy quark masses, spins and color flows are treated properly
inside {\sc alpgen}.  Heavy flavor fractions calculated using {\sc alpgen} are
calibrated against fractions measured from jet data.

The total $W$ + heavy flavor contribution is estimated by multiplying
the number of pretag $W$ + jets events in data, given in
Table~\ref{tab:pretag}, by the calculated $W$ + heavy flavor fraction
and the tagging efficiency in Monte Carlo (including the SecVtx
efficiency scale factor between data and Monte Carlo).  Because the
event tagging efficiency depends on the number of heavy flavor jets in
the fiducial region $|\eta| < 2.4$, we calculate results separately
for the case of 1 and 2 heavy flavor jets.

\subsection{\label{sec:alpgenmatching}Heavy Flavor Monte Carlo Samples}

Parton-level events from the {\sc alpgen} matrix element calculation
are fed to the {\sc herwig} parton shower program which generates
additional jets from gluon radiation.  The matrix element gives a good
description of the production of a few, widely separated partons,
whereas parton showers are better suited to model the emission of soft
collinear gluons.  Following a matrix element calculation with a
parton showering algorithm provides a better model of the data than
does either approach separately.

One outstanding issue for such a combined approach is how to avoid
double counting in the region of phase space populated both by higher
order matrix elements and the parton shower.  Specifically, the
radiation from the parton shower in a $W$ + $n$ parton Monte Carlo
sample can produce jets which cover part of the phase space
described by the $W$ + $(n+1)$ parton Monte Carlo.  Although a
rigorous combination prescription has been proposed to avoid such
double counting, it has not yet been fully implemented in any of the
matrix element Monte Carlo programs~\cite{ckkwee,ckkwhadron}.

A simple procedure deals with the possible double counting by matching
final state partons to reconstructed jets and rejecting events where
the showering algorithm has produced a hard
parton~\cite{mlmmatching,mrennamatching}.  Events are rejected if
there are extra jets which fail to match to the light partons
generated at the matrix element level or if there are missing jets.
In the special case of heavy flavor partons, the strict matching
criteria are relaxed because two partons may be merged into one jet
due to the parton mass.  Although it minimizes double counting of
generated events, this procedure introduces a new type of systematic
uncertainty which depends on the matching criteria and the jet
definition.



The matching algorithm is applied at the stable generated particle
level, before any detector simulation.  Stable particles after the
parton shower are required to have $p_T>0.4(0.0)\,\rm{GeV}/c$ for
charged (neutral) particles and $|\eta|<3$.  The jet clustering is a
simple cone clustering scheme where the number of final jets
(particles) is reduced by joining the two closest jets (particles)
within a cone of radius $\Delta R = 0.4$ into one.  Once all possible
merging is completed, the jet four-momentum is recalculated using all
of the particles inside the jet cone.  A stable-particle jet is
required to have $E_T>10\,\rm{GeV}$ and $|\eta|<2.4$, and the matched
parton must fall within a cone radius of 0.4.

The following requirements reduce event double counting after the
parton shower: 1) reject events in which an extra jet failed to match any
parton from the matrix element calculation,
2) ignore matching requirement for heavy flavor partons because
the effect of their masses has been included in the matrix element
calculation, and
3) keep only the events which pass the strict jet-light parton matching.

\begin{figure}
  \includegraphics[width=0.5\textwidth]{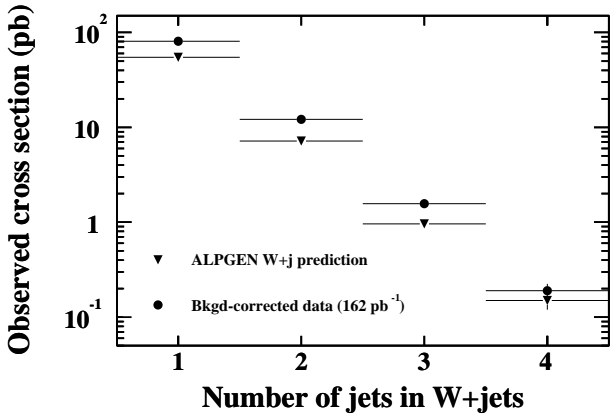}%
  \caption{\label{fig:elejets_observed}Observed $W$ + jets cross section
  compared with the {\sc alpgen} $W$ + jets prediction as a function of number
  of jets.  (Only statistical errors are shown, and the results are
  not corrected for acceptance.)}
\end{figure}

Fully exclusive matched events in each matrix element Monte Carlo
sample are summed, weighting by the appropriate cross sections.
Because the double-counted events have been removed by the matching
procedure, this combined sample should reproduce the $W$ + jets data.
These results are stable in terms of different matching algorithms,
cone size, and jet $E_T$ requirement.  The predicted $W$ + jets cross
section, without any acceptance correction, is plotted in
Figure~\ref{fig:elejets_observed} with the measurement in the electron
and muon channels.  The non-$W$ and diboson backgrounds as well as the
expected contribution from $t\bar t$ production are subtracted for
this measurement.  Even though the overall normalization of the Monte
Carlo does not reproduce the data very well, the jet multiplicity
dependences in data and Monte Carlo are in good agreement.

\subsection{\label{sec:hfmontecarlo}Heavy Flavor
  Fraction in Simulated $W$ + jets Events}

The heavy flavor fractions for $W$ + jets events, computed using an
{\sc alpgen}/{\sc herwig} Monte Carlo sample, are defined to be the
ratio of the observed $W$ + heavy flavor and $W$ + jets cross
sections.

The matching algorithm operates with particle-level jets, but jets
from a full calorimeter simulation provide better agreement with jets
in data.  A detector-level jet is required to have $E_T>15\,\rm{GeV}$
and $|\eta|<2$, and a heavy flavor jet is required to match to any $b$
or $c$ parton inside a cone with $\Delta R = 0.4$.


A summary of systematic uncertainties inherent in this heavy flavor
fraction measurement is presented in Table~\ref{tab:hfsys}.  The
matching uncertainty is estimated by recomputing the heavy flavor
fraction after varying the matching cone sizes (0.4, 0.7) and $E_T$
from 10 to 15 GeV.  We take half of the difference in the 4-jet bin as
the matching systematic uncertainty.  Uncertainties due to the
interaction energy scale $Q^2$, PDFs, and heavy quark masses are
calculated by comparing the ratio of the $Wb\bar b$ + 1 parton and $W$
+ 3 partons cross sections from {\sc alpgen} and estimating the
variation by changing the $Q^2$ (between 2$m_W^2$ and 0.5$m_W^2$),
parton distribution functions (among the 20 eigenvector pairs from
CTEQ6M\cite{stumppdf}), and the heavy quark mass ($\pm0.3\,\rm{GeV}$).
The relative systematic uncertainties in Table~\ref{tab:hfsys} are
applied to all jet multiplicity bins. The final measured heavy flavor
fractions for $W$ + jets events can be found in
Table~\ref{tab:hffracs_scaled}.

\begin{table}
  \caption{\label{tab:hfsys}Summary of systematic uncertainties in the heavy
    flavor fraction determination.}
  \begin{ruledtabular}
    \begin{tabular}{cccc}
      Source & \multicolumn{3}{c}{Uncertainty} \\
      Fractions & $Wb\bar b$ & $Wc\bar c$ & $Wc$ \\ \hline
      Matching criteria &
      15\% & 15\% & 10\% \\
      $Q^2$ scale ($2M_W^2$ to $0.5M_W^2$) & 4\% & 4\% &
      5\% \\
      PDF & 5\% & 5\% & 10\% \\
      Jet energy scale & 5\% & 5\% & 10\% \\
      ISR/FSR & 10\% & 10\% & 10\% \\
      $b,c$ masses $(4.75, 1.55\pm0.3\,\mathrm{GeV}/c^2)$ & 6\% & 10\% \\ \hline
      Total & 21\% & 22\% & 21\% \\
    \end{tabular}
  \end{ruledtabular}
\end{table}

\subsection{\label{sec:hfdijetdata}Calibration of Heavy Flavor
  Fraction Using Jet Data}

With the current data sample and a limited number of SecVtx-tagged $W$
+ jets data events, it is difficult to verify the {\sc alpgen} heavy
flavor fractions in $W$ + jets events directly with data.
Fortunately, an inclusive jet sample, without identified $W$ bosons,
is a large related class of events whose production processes are
described by Feynman diagrams similar to those of $W$ + jets events.
In particular, gluon splitting to heavy quark pairs accounts for part
of the heavy flavor production in both samples.
The inclusive QCD jet sample can be used to compare the heavy flavor
fractions calculated in Monte Carlo with results from data.
Any discrepancy between heavy flavor fractions in data and Monte Carlo
could then be used to adjust the calculated heavy flavor fractions in
$W$ + jets events.

Heavy flavor fractions are calculated in both {\sc pythia} and
{\sc alpgen}+{\sc herwig} Monte Carlo jet samples.
Events are required to have 2 or 3 jets with $E_T>15$ GeV and
$|\eta|<2.0$ and at least one jet with $E_T>20$ GeV to satisfy trigger
requirements.  Events from the {\sc alpgen} sample must also pass the
matching algorithm described in Section~\ref{sec:alpgenmatching}.  


Contributions to the jet data sample from heavy and light partons are
determined by fitting the pseudo-$c\tau$ distribution for tagged jets,
thereby discriminating between jets from $b$, $c$, and light partons
or gluons on a statistical basis.  Pseudo-$c\tau$ is defined as
$L_{\rm{2D}} \times M_{\rm{vtx}} / p_T^{\rm{vtx}}$, where
$M_{\rm{vtx}}$ is the invariant mass of all tracks in the secondary
vertex and $p_T^{\rm{vtx}}$ is the transverse momentum of the
secondary vertex four-vector.  Even though the $L_{\rm{2D}}$
distribution is similar for $b$ and $c$ quarks, the pseudo-$c\tau$ is
very different for the two flavors.

\begin{figure}
  \includegraphics[width=0.5\textwidth]{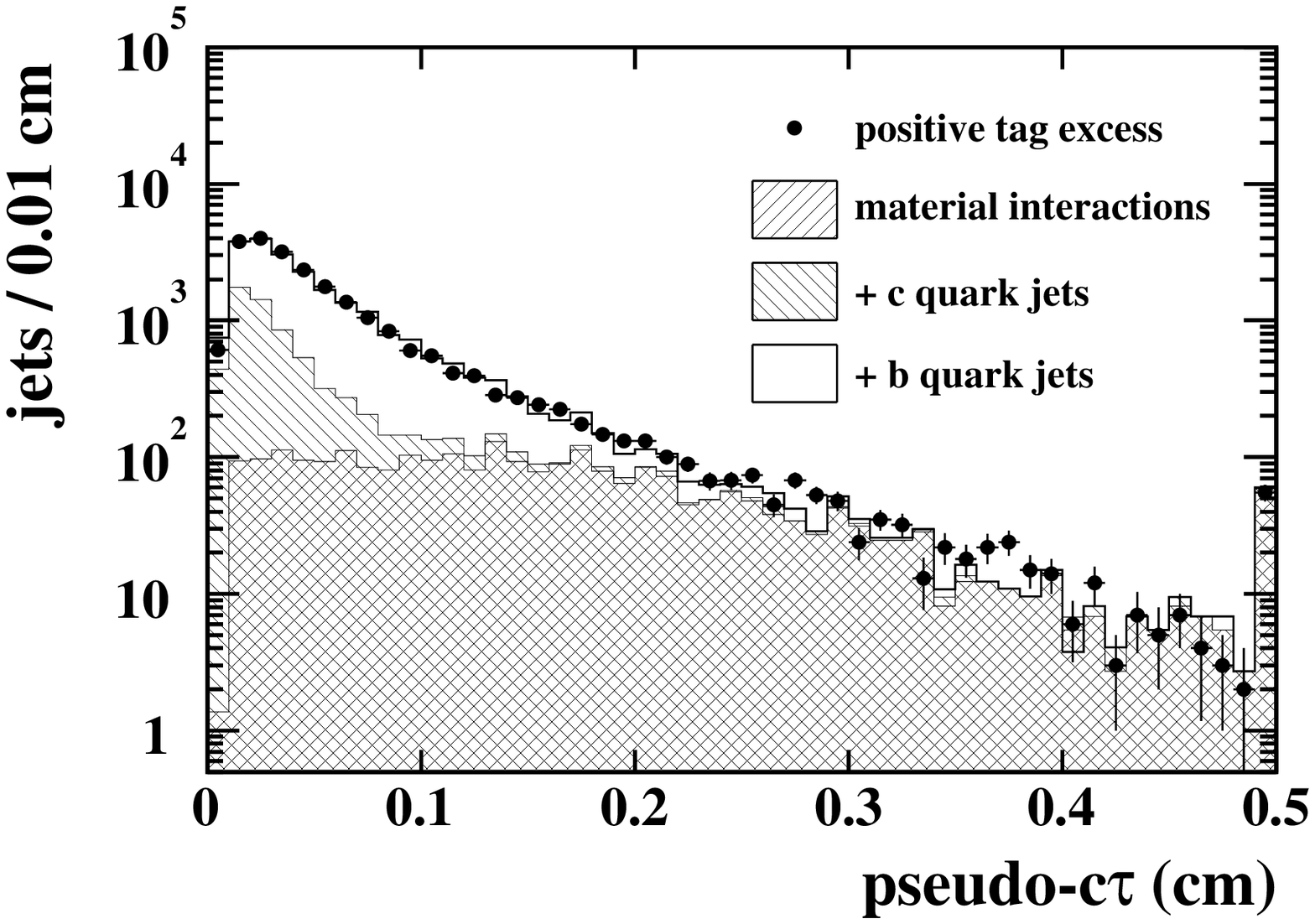}%
  \caption{\label{fig:ctau_fit}Pseudo-$c\tau$ distribution for jet data,
    including fitted contributions for the different components of
    heavy flavor and secondary interactions in light flavor jets.}
\end{figure}

The fit is made more robust by subtracting the contribution from
negative SecVtx tags and fitting the difference only, as shown in
Fig.~\ref{fig:ctau_fit}.  Template distributions of the pseudo-$c\tau$
for $b$ and $c$ jets are
derived by matching jets to partons in Monte Carlo, and a separate
template is created for secondary interactions in light quark jets,
including material interactions and long-lived $\Lambda$ and $K_s^0$
particles.

If the signed decay length distribution of secondary vertices in light
flavor jets and from tracking combinatorics were symmetric about zero,
then the number of fake positive tags from light flavor could be
simply estimated by counting the number of negative tags.
Unfortunately, secondary vertices from material interactions or
long-lived light flavor particles are more likely to have positive
decay lengths than negative decay lengths, and there are some real
heavy flavor jets with negative decay lengths.  The heavy flavor
contribution with negative decay lengths is first estimated from Monte
Carlo, and then scaled by a factor of $1.6\pm0.3$ to account for a
larger overall observed negative tag contribution in data than in
Monte Carlo events.

The net excess of secondary interactions on the positive side, $\Delta
N$, is computed from the secondary contribution fit results, after
subtracting the heavy flavor contributions on the negative side.  The
resulting average correction factor $\Delta N/N$ needed to scale the
number of negative tags to obtain the correct number of fake positive
tags is $1.2 \pm 0.1$.  This average factor is applied uniformly to
all jets, independently of jet $E_T$ and other jet properties.  The
uncertainty on this factor is due to the uncertainties in the fit
templates and the difference in $\Delta N/N$ between the different jet
$E_T$ bins.

\begin{table*}
  \caption{\label{tab:jet20fit}
    Fitted contributions from $b$, $c$ jets and 
    secondary interactions or long-lived light flavor particles in data
    events.  The uncertainties on the $b$ and $c$ fractions are total
    uncertainties including 5\% and 10\% uncertainties due to the
    templates.  The ratio $\Delta N / N$ estimates the excess of
    positive over negative tags in data events, due to secondary
    interactions and long-lived light flavor particles.} 
  \begin{ruledtabular}
    \begin{tabular}{cccccc}
   $E_T$ (GeV)      & $E_T<25$ & $25<E_T<35$ & $35<E_T<45$ & $E_T>45$ & All \\ \hline 
   Taggable   & 858,643 & 415,373 & 128,994 & 77,632 & 1,480,642\\ 
   Pos. - Neg.& 12,208  & 7131  & 2511 & 1596 & 23,446\\
   Negative   & 3283   & 1999  &  803 &  697 & 6782 \\
   Fitted $b$s & $7937\pm 483$ & $4412\pm 312$ & $1609\pm 131$ &
   $843\pm 102$ & $15,147\pm507$\\
   Fitted $c$s & $3040\pm 427$ & $1858\pm 276$ & $520\pm 110$ & $407\pm
   93$ & $5589\pm451$\\
   Secondary  & $1284\pm 142$ & $900\pm 102$  & $379\pm 50$  & $324\pm
   39$ & $2836\pm171$ \\ \hline 
   $\Delta N$ & $482\pm 224$ & $431 \pm 144$ & $230\pm 59$ & $227\pm
   44$ & $1336\pm365$\\
   $\Delta N / N (\%)$ & $15\pm 7$ & $22 \pm 7$ & $29 \pm 7$ & $32 \pm
   7$ & $20\pm5$\\
   $b$s/Jets (\%) & $0.92\pm 0.08$ & $1.06\pm 0.10$ &$1.25\pm 0.12$ &
   $1.09\pm 0.14$ & $1.02\pm0.06$ \\ 
   $c$s/Jets (\%) & $0.35\pm 0.06$ & $0.45\pm 0.08$ & $0.40\pm 0.10$ &
   $0.52\pm 0.13$ & $0.38\pm0.05$ \\
   \end{tabular} 
  \end{ruledtabular}
\end{table*} 

The heavy flavor fraction as a function of jet $E_T$ is stable, as
shown in Table~\ref{tab:jet20fit}, where an uncertainty of 5\% (10\%)
for the $b$ ($c$) fraction is included due to template uncertainties.
These results include the effect of the efficiency scale factor
between data and simulation.  Measured heavy flavor fractions from the
data are consistently 50\% higher than the {\sc alpgen} prediction,
for both $b$ and $c$ jets, although the {\sc pythia} calculation seems
to match the data more closely.  These heavy flavor fractions are
compared with the heavy flavor fractions calculated using {\sc alpgen}
inclusive jet Monte Carlo with the matching prescription.  On average
the data/{\sc alpgen} ratio is $1.5 \pm 0.4$, where the uncertainty is
dominated by the systematic uncertainties associated with the {\sc
  alpgen} heavy flavor calculations (Table~\ref{tab:hfsys}).  From
these fits alone, it is not clear if the discrepancy is consistent for
all production diagrams or only for some subsets of gluon splitting to
heavy flavor partons.

\begin{figure}
  \includegraphics[width=0.5\textwidth,clip=true]{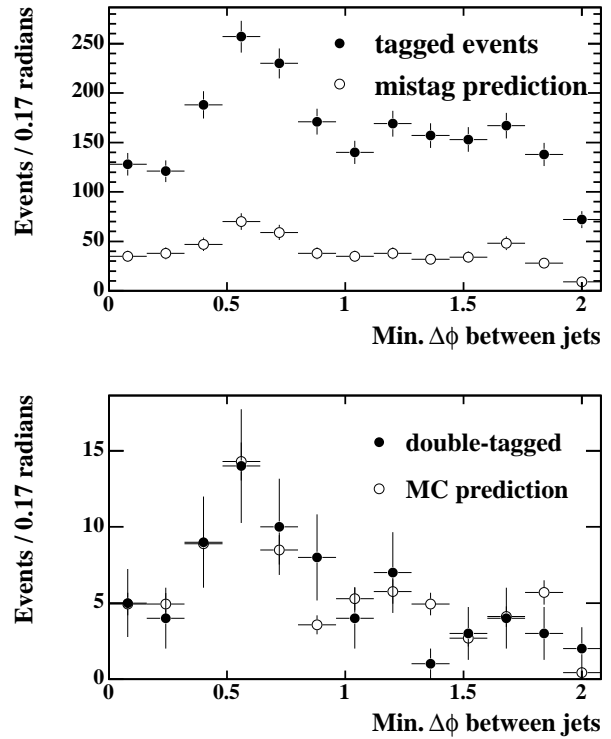}%
  \caption{\label{fig:dphi_3jet}Distribution of closest
    jets in $\Delta \phi$ for tagged 3-jet data events with fake tag
    prediction (top) and for double-tagged events in which the
    tagged jets are also the two closest jets (bottom). }
\end{figure}

Because jets with gluon splitting have a small opening angle, the
distribution of $\Delta \phi$ between the two closest jets in an event
highlights the contribution from gluon splitting.  A sample of events
with 2 tagged jets is selected from the 3-jet sample and compared to
Monte Carlo.  The mistag contribution is removed from the
double-tagged samples by subtracting events with one or more negative
tag.  The good agreement, shown in Fig.~\ref{fig:dphi_3jet}, indicates
that the gluon splitting contribution relative to other production
mechanisms is well-modelled.

\begin{figure}
  \includegraphics[width=0.5\textwidth]{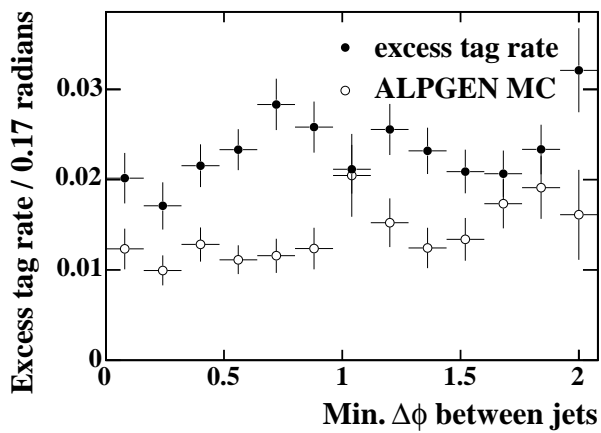}%
  \caption{\label{fig:dphi_3jet_rate}The positive tag
    excess rate in data and {\sc alpgen} Monte Carlo as a function of $\Delta
    \phi$.}
\end{figure}

Another sample with gluon splitting contributions, this time of
single-tagged 3-jet events, can be used to check the dependence of the
data/{\sc alpgen} normalization factor.  When the excess tag rate,
interpreted as the heavy flavor fraction, is plotted as a function of
minimum $\Delta \phi$ between jets (Fig.~\ref{fig:dphi_3jet_rate}),
there is no evidence of structure in the fractions as a function of
$\Delta \phi$ even though the heavy flavor fractions in data are still
1.5 times the heavy flavor fractions in Monte Carlo.  This consistency
disfavors the hypothesis of missing or under-represented heavy flavor
production diagrams.

The measured ratio of $1.5\pm0.4$ between the heavy flavor fractions
in the {\sc alpgen}/{\sc herwig} samples and the data is not
inconsistent with other recent studies, which indicate that a
$K$-factor may be necessary to account for higher-order
effects~\cite{method2nlo}.  Based on this calibration with the jet
data sample, we scale the expected $Wb\bar b$ and $Wc\bar c$
background contributions derived from {\sc alpgen} by a factor of
$1.5\pm0.4$.  Since the $Wc$ background is produced through a different
diagram, that contribution is not rescaled.

Table~\ref{tab:hffracs_scaled} summarizes the one and two $b$ ($c$)
fractions as a function of jet multiplicity, as well as the
corresponding SecVtx tagging efficiencies, where the efficiency scale
factor, as calculated in Section~\ref{sec:secvtx}, has already been
applied.  The 1B (1C) fractions are for events with exactly one jet
matched to a $b$ ($c$) parton, and the 2B (2C) fractions are for events
with exactly two jets matched to $b$ ($c$) partons.  These values are used
in Section~\ref{sec:wflavor} to predict the background contribution
from $W$ + heavy flavor production

\begin{table*}
  \caption{\label{tab:hffracs_scaled}Ratio of $W$ + heavy flavor production
    to total $W$ + jet production, for different jet multiplicities.
    The heavy flavor ratios include the correction factor $1.5\pm0.4$
    as measured from jet data, and the SecVtx event tagging
    efficiencies include the scale factor described in
    Section~\ref{sec:secvtx}.  These values are used in
    Section~\ref{sec:wflavor} to predict the background contribution from $W$ +
    heavy flavor production.}
  \begin{ruledtabular}
    \begin{tabular}{ccccccc}
      Jet multiplicity & 1 jet & 2 jets & \multicolumn{2}{c}{3 jets} &
      \multicolumn{2}{c}{$\ge$4 jets} \\
      $H_T$ (GeV)& & & $H_T > 0$ & $H_T > 200$ & $H_T > 0$ & $H_T > 200$ \\ \hline
      \multicolumn{7}{c} {W + HF fractions (\%)} \\ \hline
      1B &  $1.0\pm 0.3$ &   $1.4\pm 0.4$  & $2.0\pm 0.5$ &
      $2.4\pm 0.6$ & $2.2\pm 0.6$ & $2.2\pm 0.6$ \\
      2B & & $1.4\pm 0.4$ &  $2.0\pm 0.5$ &  $2.3\pm 0.6$ &
      $2.6\pm 0.7$ & $2.6\pm 0.7$ \\
      1C & $1.6\pm 0.4$ & $2.4\pm 0.6$ & $3.4\pm 0.9$ & $3.8\pm
      1.0$ & $3.6\pm 1.0$ &  $3.5\pm 1.0$ \\   
      2C & & $1.8\pm 0.5$ & $2.7\pm 0.7$ & $2.9\pm 0.8$ & $3.7\pm 1.0$
      & $3.7\pm 1.0$ \\
      Wc & $4.3 \pm 0.9$ & $6.0 \pm 1.3$ & $6.3 \pm 1.3$ & $6.0 \pm
      1.3$ & $6.1 \pm 1.3$ & $5.9 \pm 1.3$ \\ \hline
      \multicolumn{7}{c} {SecVtx tagging efficiencies (\%)} \\ \hline 
      1B($\ge 1$tag) & $26.8\pm 2.0 $ & $27.8\pm 2.2$ & $29.3\pm 2.5$
      & $30.9 \pm 2.9$ & $24.2\pm 3.3$ & $27.4\pm 3.8$ \\ 
      2B($\ge 1$tag) &  & $48.6\pm 3.2 $ & $50.0\pm 3.8$ & $52.6\pm
      4.5$ & $50.3\pm 4.9$ & $50.0\pm 5.1$ \\ 
      2B($\ge 2$tags) &  & $9.1\pm 1.4$ & $9.5\pm 1.5$ & $10.4 \pm
      1.6$ & $8.1\pm 1.4$ & $8.6 \pm 1.5$ \\ 
      1C($\ge 1$tag) & $6.2\pm 0.9 $ & $6.7\pm 1.0$ & $6.1\pm 1.1$ &
      $6.6\pm 1.3$ & $7.7\pm 1.9$ & $7.5\pm 2.0$ \\ 
      2C($\ge 1$tag) &  & $12.3\pm 1.9 $ & $11.6\pm 2.0$ & $12.6\pm
      2.5$ & $10.1\pm 2.3$ & $9.6\pm 2.4$ \\
      2C($\ge 2$tags) & & $0.5 \pm 0.2$ & $0.4 \pm 0.1$ & $0.5 \pm
      0.2$ & $0.8 \pm 0.4$ & $0.9 \pm 0.4$ \\   
      Wc ($\ge 1$tag) & $5.8\pm 0.9$ & $6.1\pm 0.9 $ & $7.1\pm 1.2$ &
      $7.6\pm 1.5$ & $5.6\pm 1.6$ & $5.8\pm 1.8$ \\ \hline
    \end{tabular}
  \end{ruledtabular}
\end{table*}

%% file: backgrounds.tex
\section{\label{sec:m2bkg}Backgrounds in the Tagged $W$ + jets Sample}

The non-$t\bar t$ events in the $W$ + jets sample are from
direct QCD production of heavy flavor without an associated $W$ boson,
mistags of light quark jets in $W$ + jets events, $W$ + heavy
flavor production, and other low rate electroweak processes with heavy
flavor such as diboson and single top production.  The estimation of
each of these backgrounds is described in turn.

\subsection{\label{sec:qcd}Non-$W$ QCD Background}

The non-$W$ QCD background is a mixture of events where the lepton does
not come from the decay of a $W$ or $Z$ boson.  These include lepton and
missing energy fakes as well as semileptonic $b$ hadron decays. Since
several backgrounds are calculated by normalizing to the number of $W$ +
jets events before tagging, it is necessary to understand the level of
QCD contamination in the pretag sample. In addition, some of these
non-$W$ QCD events may be $b$-tagged.  Both the pretag and tagged
contributions are measured directly from data events.


In a leptonic $W$ decay, the lepton is isolated and there is large
$\met$ due to the neutrino, while in non-$W$ events this is not
necessarily true. We define the lepton isolation, $\cal I_{\rm sol}$, as the
ratio of energy (not due to the lepton) in the calorimeter in a cone
around the lepton direction to the measured electron (muon) energy
(momentum). Isolated leptons will have small values of $\cal
I_{\rm sol}$. Sideband regions for lepton isolation and $\met$ in the
high-$p_T$ lepton sample contain mostly non-$W$ events and are used to
extrapolate QCD expectations in the signal region. The sideband
regions are defined as follows:

\begin{enumerate}
\item
Region A: $\cal I_{\rm sol}$ $>$ 0.2 and \met $<$ 15 GeV
\item
Region B: $\cal I_{\rm sol}$ $<$ 0.1 and \met $<$ 15 GeV
\item
Region C: $\cal I_{\rm sol}$ $>$ 0.2 and \met $>$ 20 GeV
\item
Region D ($W$ signal region): $\cal I_{\rm sol}$ $<$ 0.1 and \met $>$ 20 GeV
\end{enumerate}
                    
For the QCD background these two variables are assumed to be 
mostly uncorrelated: the ratio
of non-$W$ events at low and high $\cal I_{\rm sol}$ values in the low $\met$ region
is the same as in the high $\met$ region. The number of non-$W$ events
in the signal region is estimated by

\begin{equation}
QCD_{D} = \frac{N_B \times N_C}{N_A}.
\label{eqn:nonw}
\end{equation}


The contribution of true $W$ and $t\bar t$ events in the sideband regions
is estimated using Monte Carlo samples to determine the ratio of $W$ and
$t\bar t$ in the signal and sideband regions, and normalized to the
observed number of events in the pretag signal region. The correction is
5-30\% depending on the lepton type and event jet multiplicity.


\subsubsection{\label{sec:pretagqcd}Pretag Backgrounds}

The non-$W$ QCD background is calculated separately for the electron and
muon channels, as well as for different jet multiplicities.
Table~\ref{tab:finalqcd} gives the predicted QCD background fraction in
the signal region.  The main source of systematic uncertainty is the
underlying assumption that the lepton isolation and $\met$ are
uncorrelated for this background.  A study of non-isolated leptons
indicates that this assumption adds a 25\% systematic uncertainty to
the non-$W$ QCD background estimate.


\subsubsection{\label{sec:tagqcd}Tagged Backgrounds}

Some of the non-$W$ QCD events are $b$-tagged and end up in the final
event count. One estimate of this contribution applies
Equation~\ref{eqn:nonw} to the tagged event sample, but this method is
limited by the tagged sample size.  To increase the number of events,
regions A and C are redefined by lowering the isolation boundary to
the edge of the signal region, $\cal I_{\rm sol}$ $> 0.1$. The precision on
this estimate is limited by the number of tagged events in the
sideband regions.

A second method scales the pretag QCD fraction by the average tagging
rate for QCD events.  This method has the advantage of normalizing the
background with the larger statistics of the pretag sample, but
requires a reliable estimate of the tag rate.  The tagging rate in
region B for events with two or more jets is applied to the number of
taggable jets in the signal region times the pretag QCD background
fraction.  

Both background estimates contribute to the weighted average shown in
Table~\ref{tab:finalqcd}.


\begin{table*}[hbtp]
\caption{\label{tab:finalqcd} Non-$W$ QCD background estimate.  Results from
  the tag rate method and the tag sample method are the number of
  events expected in the b-tagged lepton + jets sample.}
\begin{ruledtabular} 
\begin{tabular}{ccccccc}
& & & \multicolumn{2}{c}{ $H_T >$ 0 } & \multicolumn{2}{c}{ $H_T >$ 200 GeV } \\
Jet multiplicity & 1 jet & 2 jets & 3 jets & $\geq 4$ jets &3 jets & $\geq 4$ jets  \\ \hline 
\multicolumn{7}{c}{ Electrons}\\
\hline
Pretag non-$W$ QCD Fraction   &  0.14 $\pm$ 0.04 & 0.17 $\pm$ 0.04 & \multicolumn{4}{c}{ 0.20 $\pm$ 0.05 } \\ \hline
Tag Rate Method & 16.3 $\pm$  4.7 & 7.4 $\pm$ 2.2 & \multicolumn{2}{c}{ 3.2 $\pm$ 1.0 }  & \multicolumn{2}{c}{ 2.1 $\pm$ 0.6 } \\
Tag Sample Method &   21.8$\pm$ 3.8 &  10.0$\pm$ 2.2 &  \multicolumn{2}{c}{ 4.9 $\pm$ 1.3 } & \multicolumn{2}{c}{ 2.6 $\pm$ 0.8 } \\
Combined Tag Estimate & 19.6 $\pm$ 3.0 & 8.7 $\pm$ 1.6 & 2.7 $\pm$ 0.6 & 1.1 $\pm$ 0.2 & 1.3 $\pm$ 0.3 & 1.0 $\pm$ 0.3 \\
\hline
\hline
\multicolumn{7}{c}{ Muons }\\
\hline
Pretag non-$W$ QCD Fraction & 0.034 $\pm$ 0.010 & 0.043 $\pm$ 0.011 & \multicolumn{4}{c}{0.075 $\pm$ 0.023} \\ \hline
Tag Rate Method & 4.0 $\pm$ 1.3 & 1.2 $\pm$ 0.6 &  \multicolumn{2}{c}{ 0.7 $\pm$ 0.3 } &  \multicolumn{2}{c}{ 0.5 $\pm$ 0.1 } \\
Tag Sample Method &  4.8$\pm$ 1.1 &  1.8$\pm$ 0.5 &  \multicolumn{2}{c}{ 1.3 $\pm$ 0.4 } &  \multicolumn{2}{c}{ 1.0 $\pm$ 0.3 }  \\ 
Combined Tag Estimate & 4.5 $\pm$ 0.8 & 1.5 $\pm$ 0.4 & 0.7 $\pm$ 0.2 & 0.2 $\pm$ 0.1 & 0.3 $\pm$ 0.1 & 0.3 $\pm$ 0.1       \\
\hline\hline
Electron+Muon & 24.3$\pm$ 3.5 &  10.5$\pm$ 1.9 &  3.4$\pm$ 0.7 &  1.3$\pm$ 0.3 &  1.6$\pm$ 0.4 &  1.2$\pm$ 0.4 \\
\end{tabular}
\end{ruledtabular}
\end{table*} 

\subsection{\label{sec:mistag}Mistags} 

\begin{figure}[htbp]
\includegraphics[width=0.5\textwidth]{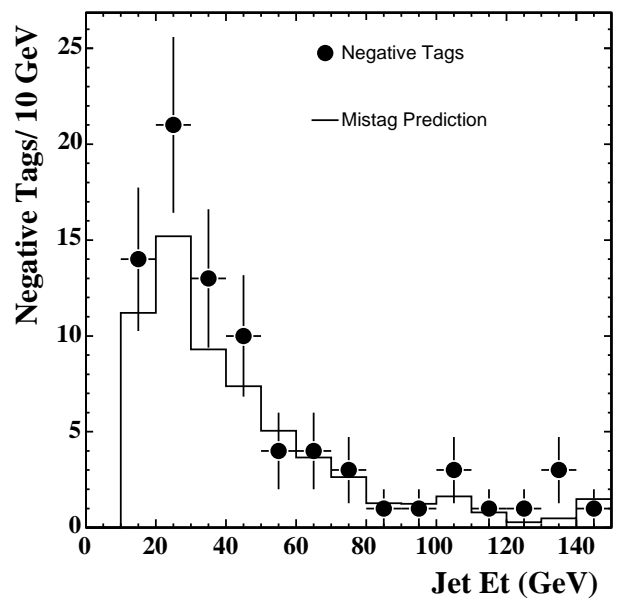}
\caption{\label{fig:negtag_jetet}Comparison of observed and predicted 
negative SecVtx tags {\em vs.} jet $E_T$ in the lepton + jets sample.}
\end{figure}

Mistag background events are $W$ + jets events where the tagged jet does
not result from the decay of a heavy quark.
As described in Sec.~\ref{sec:mistags}, the mistag
rate per jet is parameterized as a function of the number of
tracks, the raw jet $E_T$, the $\eta$ and $\phi$ of the jet, and the
sum of the $E_T$ for all jets in the event with $E_T > 10$ GeV and
$|\eta | < 2.4$.  To estimate the size of the mistag background, 
each jet is weighted with its mistag rate in the pretag sample.  The
sum of the weights over all jets in the sample is then scaled down by the fraction
of pretag events which are due to QCD background, as in
Sec.~\ref{sec:pretagqcd}, since these have already been counted in the
procedure of Sec.~\ref{sec:tagqcd}.  
The low mistag rate per jet means that a negligible number of events
have more than one mistagged jet; therefore, the number of mistagged
jets is a good approximation of the number of events with at least one
mistagged jet.
This method is
tested by comparing the negative SecVtx tags observed and predicted
for the pretag sample as a function of the jet $E_T$, plotted in
Figure~\ref{fig:negtag_jetet}.  There is reasonable agreement in the shape
and normalization of the prediction.

For the estimate of the number of fake positive tags, the mistag
correction factor of 1.2 $\pm$ 0.1 described in
Sec.~\ref{sec:hfdijetdata} is applied to account for additional
mistags of light quark jets due to material interactions or long lived
light quark hadrons.  The final results for the mistag estimate are
shown in Table~\ref{tab:corrbkgdsingle}. The error includes
statistical uncertainties from the pretag sample, including the small
effect of correlation between mistag weights that come from the same
bin. In addition, an 11\% total systematic uncertainty includes
uncertainty due to the sample dependence of the mistag rate
parameterization and the uncertainty on the mistag correction factor
of 1.2 for the positive/negative mistag asymmetry.

\subsection{\label{sec:wflavor}$W$ + Heavy Flavor Backgrounds}

The production of $W$ bosons associated with heavy flavor in the
processes $Wb\bar b$, $Wc\bar c$, and $Wc$ is a significant part of the background
for the tagged sample.  The techniques described in
Sec.~\ref{sec:hffracs} are used to estimate the fraction of the
inclusive $W$ + jets events which have $Wb\bar b$, $Wc\bar c$, and $Wc$.  The number
of $Wb\bar b$, $Wc\bar c$, and $Wc$ events is given by multiplying the heavy flavor
fractions by the pretag event count, after subtracting the non-$W$
backgrounds.  Estimates of the tagged background are then obtained by
multiplying the tagging efficiencies summarized in
Table~\ref{tab:hffracs_scaled}.



The pretag $W$ + jets sample includes some contribution from
misidentified $Z\rightarrow\mu^+\mu^-$ events. The heavy
flavor fraction for that process is twice as large as for the $W$
events. The extra contribution of heavy flavor from $Z$ events is
described in Section~\ref{sec:checks} and given in
Table~\ref{tab:zjets}. Corrections due to $t\bar t$ contributions 
in the pretag events are discussed in Section~\ref{sec:singletags}.  

\subsection{\label{sec:mcbkg} Other Backgrounds}
 
A number of backgrounds are too small to be measured directly, thus we
use the Monte Carlo to predict their contribution to the sample. The
diboson production processes $WW$, $WZ$, and $ZZ$, in association with
jets, can mimic the $t\bar t$ signal when one boson decays
leptonically and the other decays to a taggable $b$ or $c$ quark jet.
The process $Z\rightarrow \tau^+ \tau^-$, in association with jets,
can mimic the signal when one $\tau$ decays leptonically and the other
hadronically.  Top quarks are expected to be produced singly with a
$t\bar b$ final state through $s$-channel $\qqbar$
annihilation, and $t$-channel $W$-gluon fusion processes.

We use Monte Carlo samples to measure the acceptance and tagging
efficiency.  The Monte Carlo acceptance is corrected for the lepton
identification and trigger efficiencies as is done for the $t\bar t$
acceptance as described in Section~\ref{sec:optimization}.  The
tagging efficiency is scaled by the MC/data tagging scale factor, with
double the uncertainty for tagging charm jets as in $W \rightarrow
c\bar s$.  The normalization is based on the measured integrated
luminosity and the following theoretical cross sections
$\sigma(\mathrm{single\, top}) = 2.86 \pm 0.09\,\mathrm{pb}$,
$\sigma(WW) = 13.25 \pm 0.25\,\mathrm{pb}$, $\sigma(WZ) = 3.96 \pm
0.06\,\mathrm{pb}$, and $\sigma(ZZ) = 1.58 \pm
0.02\,\mathrm{pb}$~\cite{campbell_dibosonxsec,singletop}.

\subsection{\label{sec:bkgsum}Background Summary}

\begin{figure}[hbtp]
\includegraphics[width=0.5\textwidth]{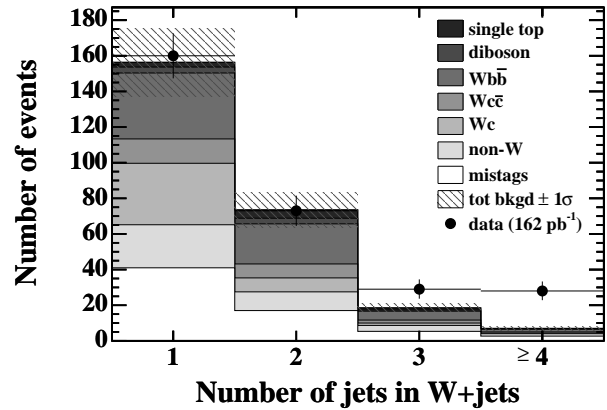}
\caption{\label{fig:plotalltags} Number of events passing the
  selection criteria with at least one tagged jet, and the background
  prediction for the same selection. The $H_T$ requirement has not
  been applied.}
\end{figure}

A complete summary of all of the background contributions is given in
Table~\ref{tab:corrbkgdsingle}. Figure~\ref{fig:plotalltags} shows the
contribution of the different backgrounds for each jet bin compared to
the number of data events satisfying all of the selection criteria and
having at least one positively tagged jet.  We find good agreement
between background and data in the one and two jet bins, validating our
background calculation. The excess of tags in the three and four jet bins is
attributed to $t\bar t$.  We have already described how the estimates for
$Wb\bar b$, $Wc\bar c$, $Wc$ ~and mistags, which depend on the number of $W$ pretag
events in the data, are corrected for the contribution of QCD
backgrounds to the pretag sample. A similar correction needs to be
made to account for the real $t\bar t$ in the pretag sample. This is
done as part of the cross section measurement as described in
Section~\ref{sec:singletags}. We find the pretag sample to be 10-15\%
$t\bar t$ in the three jet bin and 40-50\% $t\bar t$ in the four jet bin.

%% file: cross_checks.tex
\section{\label{sec:checks}Cross Check Using the $Z$ + Jets samples}

An investigation into the $Z$ + jets sample provides a good cross check
on our background calculations since the $t\bar t$ and non-$W$ QCD
contamination in these samples is small. 

%


The heavy flavor contribution in $Z$ + jets is expected to be close to
that in $W$ + jets in terms of gluon splitting.  However, there is an
additional diagram $gg \rightarrow Zb\bar b$, which is not
present in $W$ + jets. We use the same procedures described in
previous sections to estimate the heavy flavor fractions using the $Z$
+ jets {\sc alpgen} Monte Carlo samples.  The fraction of $Zc\bar c$ events
(including $Zc$) is approximately twice the fraction of $Wc\bar c$ events, and
the fraction of $Zb\bar b$ events is approximately twice the fraction of
$Wb\bar b$ events.  The heavy flavor fractions in $Z$ + jets are therefore
estimated by multiplying the above factors with the heavy flavor
fractions in $W$ + jets listed in Table~\ref{tab:hffracs_scaled}.

Events with a $Z$ boson are selected by identifying oppositely charged
$e^+e^-$ and $\mu^+\mu^-$ pairs with an invariant mass between 75 and
105 GeV/$c^2$. Both leptons are required to pass the tight lepton
selection used for the $W$ + jets analysis in order to collect a pure
sample of $Z$ candidates.

\begin{figure}[htbp]
  \begin{center}
    \includegraphics*[width=0.5\textwidth]{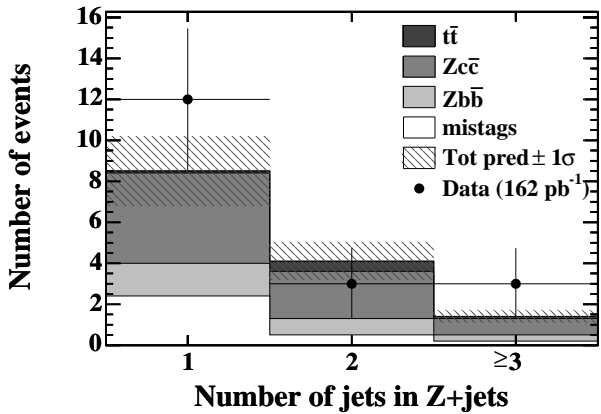}
    \caption{\label{fig:zjets1}Comparison of the observed and
      predicted number of events in the $b$-tagged $Z$ + jets sample.}
  \end{center}
\end{figure}

Table~\ref{tab:zjets} lists the yield of $Z$ candidates and the number
of tagged events observed as a function of jet multiplicity. The
background predictions are also given and are calculated in the same
way as in the previous sections for the $W$ + jets sample:
$14.0\pm1.9$ events are predicted and 18 are observed in the $Z$ +
jets sample (Fig.~\ref{fig:zjets1}).

Some $Z$ + jets events which fail the standard $Z$ removal contribute to
the $W$ + jets sample.  The fraction of $Z \rightarrow \mu^+\mu^-$
events left in the $W$ sample is about $72\pm 8\%$ of the
number of events observed in $Z \rightarrow \mu^+\mu^-$ decay.
The contribution of $Z \rightarrow e^+ e^-$, on the other
hand, is negligible.  Since those $Z$ events left in the $W$ sample have a
higher heavy flavor fraction than the $W$ events, a correction factor
accounts for the additional tagged events expected in the $W$ + jets
sample.

\begin{table} 
  \caption{\label{tab:zjets}The predicted number of $Z$ + jets events and the
    observed number, along with the $Z$ + jets contribution left in the
    $W$ + jets sample and the estimate of the resulting extra $b$ tags in
    that sample.  (The prediction of extra $b$-tagged events is included
    in the predicted background summary for the $W$ + jets sample.)}
  \begin{ruledtabular}
    \begin{tabular}{cccc}
      Jet multiplicity & $Z$+1 jet & $Z$+2 jets & $Z$+$\ge$3 jets \\ \hline
      $Z\rightarrow e^+ e^-$ & 410  & 48 & 10  \\ 
      $Z\rightarrow \mu^+\mu^-$ & 402 &59  &15   \\
      $Z\rightarrow \ell^+\ell^-$ & 812 & 107  & 25 \\ \hline 
      Mistags    & $2.4\pm 0.2$  & $0.49\pm 0.06$   & $0.23\pm 0.04$    \\
      $Zb\bar b$ & $1.6\pm 0.4$ &  $0.8\pm 0.2$  & $0.26\pm 0.08$     \\  
      $Zc\bar c$ & $4.4\pm 1.3$  & $2.3\pm 0.7$  & $0.8\pm 0.2$  \\  
      top ($\sigma_{t\bar t} = 5.6\pm 1.4$)         & $0.08\pm 0.02$ &
      $0.5\pm 0.1$ & $0.13\pm 0.03$  \\ \hline
      Pred. Total  & $8.5\pm 1.7$ & $4.1\pm 0.9$   & $1.4\pm 0.3$   \\
      Observed Events& 12 &  3  &  3    \\ \hline  
      Pretag $W$+jets & $289\pm 35$ & $42\pm 7$ & $11\pm 3$  \\ 
      Tagged in $W$+jets  & $1.1\pm 0.3$ & $0.6\pm 0.2$ & $0.2\pm 0.1$ \\
    \end{tabular}
  \end{ruledtabular}
\end{table}
 

%% file: optimization.tex
\section{\label{sec:optimization}Event Selection Optimization and Acceptance}


The event selection described in Section~\ref{sec:data}, combined with
the requirement that at least one jet be positively $b$-tagged, yields
a clean sample of top decays in the lepton plus jets channel; the
expected signal over background ratio is of the order of 2:1.  Several
ways of optimizing the event selection were studied in order to
maximize the significance of the cross section measurement, and an
inclusive event variable was found to have the greatest power to
discriminate $t\bar t$ signal from background events.  The following
section discusses event selection optimization using the $H_T$
variable.

\subsection{\label{subsec:optimization2}Optimization with the $H_T$ variable.}

The event quantity $H_T$ is defined as the scalar sum of the transverse
energy of all the kinematic objects in the event (transverse momentum
for muons), including all jets with $E_T>8$ GeV and $|\eta|<2.5$:

$$H_T = \Sigma_\mathrm{all\ jets} E_T + \met + E_T^\mathrm{electron}
\ \mathrm{or} \ p_T^\mathrm{muon}$$
 
\begin{figure}
\includegraphics*[width=0.5\textwidth,clip=true]{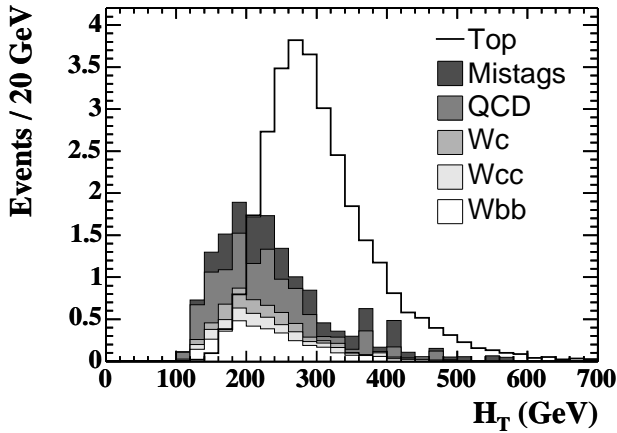}
\caption{\label{fig:htbgd}Distribution of the $H_T$ variable for
  $t\bar t$ Monte Carlo, and for various backgrounds normalized to an
  integrated luminosity of $107\,\mathrm{pb}^{-1}$.}
\end{figure}

Because of the large mass of the top quark, $H_T$, which is
representative of the hard scatter of the event, tends to be
significantly larger for $t\bar t$ events than for the backgrounds.
Figure~\ref{fig:htbgd} shows the distribution of the $H_T$ variable
after all selection cuts have been applied, including $b$-tagging, for
$t\bar t$ Monte Carlo, and for the main backgrounds: $W$ + heavy
flavor, non-$W$ QCD, and mistags.  The $W$ + heavy flavor distributions
are taken from {\sc alpgen} Monte Carlo, but all other background
shapes are estimated from data; the features visible at high $H_T$
in the non-W QCD and mistags distributions are due to poor
statistics in the control samples.

The $H_T$ distributions for the three major backgrounds are estimated
using methods described in Section~\ref{sec:m2bkg}.  The {\sc alpgen} Monte
Carlo generator is used to estimate the shape of the distribution for
the $W$ + heavy flavor background.  The non-$W$ QCD background shape is
evaluated by selecting pretag events where the lepton is not
isolated (isolation $>0.2$), while all other kinematic cuts remain
unchanged.  This subsample is presumably dominated by QCD events with
kinematic properties identical to the QCD background events that
satisfy the event selection (isolation$<0.1$). Each event in the
sub-sample is then weighted by the total positive tagging rate
measured from the jet sample (see Section~\ref{sec:mistag}).  The
mistag background shape is estimated from the pretag sample, where
each event is weighted by the negative tag rate measured from the
jet sample.  Other backgrounds (which account for less than 10\% of
the total background) are included in the overall normalization, with
the implicit assumption that their shape is not significantly
different from the others.  The $t\bar t$ contribution is normalized to
the theoretical cross section.

\begin{figure}
\includegraphics*[width=0.5\textwidth,clip=true]{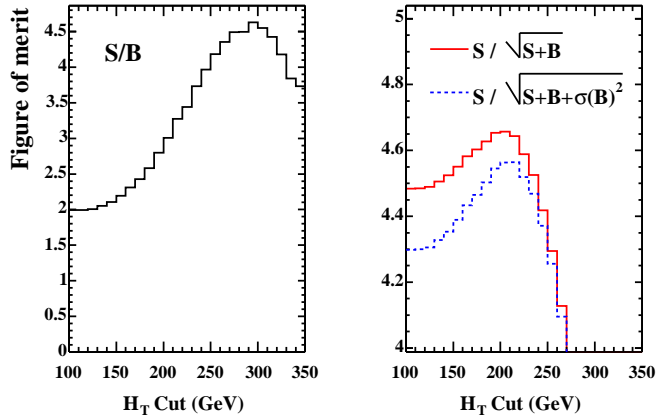}
\caption{\label{fig:htsb2} Estimate of S/B, statistical and total
  significance for $107\,\mathrm{pb}^{-1}$ integrated luminosity, as a
  function of $H_T$ cut.}
\end{figure}

Figure~\ref{fig:htbgd} shows that signal and background can be
separated by the use of the $H_T$ variable.  Figure~\ref{fig:htsb2}
shows the signal over background ratio and cross section sensitivity
as a function of an $H_T$ cut, computed from figure~\ref{fig:htbgd}.
The statistical sensitivity ($S/\sqrt{S+B}$) is compared to the total
sensitivity ($S/\sqrt{S+B+\sigma(B)^2}$, where $\sigma(B)$ is the
absolute systematic error on the background estimate). Systematic
uncertainties arising from the $H_T$ cut itself are described in
Section~\ref{sec:singletags}; they are small enough to be neglected in
the optimization process.  A cut requiring $H_T>200$ GeV is found to
be optimal: such a cut keeps 96\% of the signal and rejects 39\% of
the background; this improves the signal over background ratio from 2
to 3 and the total significance on the $t\bar t$ cross section
measurement by 6\%.

\subsection{Acceptance}

The acceptance is defined as the fraction of produced $t \bar t$
events that satisfy all trigger and selection criteria. It includes
trigger efficiency, reconstruction efficiency, as well as the
efficiency of the kinematic cuts, and of the $b$-tagging algorithm. The
branching ratios of the various decay modes of the $t \bar t$ pair
are included as well.

The contributions to the acceptance are divided as follows:
\begin{equation}
\label{eq:cs}
\epsilon_{t\bar t} =
\epsilon_{\mathrm{trig}} \cdot \epsilon_{z_0} \cdot \epsilon_{\mathrm{veto}} \cdot 
\mathcal{\epsilon}^{\mathrm{MC}}_{t \bar t} \cdot k_{\mathrm{lep-id}} \cdot 
\epsilon_{\mathrm{tag-event}}
\end{equation}
where $\epsilon_{z_0}$ is the efficiency of the $|z_0|\le60$cm cut,
$\mathcal{\epsilon}^\mathrm{MC}_{t \bar t}$ is the fraction of Monte Carlo
$t\bar t$ events which pass all the selection cuts (except for
$b$-tagging), $\epsilon_{\mathrm{veto}}$ is the combined efficiency of
the various vetoes (conversion removal, cosmic removal, dilepton and
$Z^0$ rejections), $\epsilon_{\mathrm{trig}}$ is the trigger
efficiency for identifying high $p_T$ leptons, and
$\epsilon_{\mathrm{tag-event}}$ is the efficiency to tag at least one
jet in a $t \bar t$ event.  $k_{\mathrm{lep-id}}$ is a factor that
corrects for the lepton identification efficiency difference between
data and Monte Carlo.

The efficiency of the $z_0$ cut is measured from data and found to be
$\epsilon_{z_0} = 0.951 \pm 0.005$.  The trigger efficiency
$\epsilon_{\mathrm{trig}}$ is different for each type of lepton
trigger: $\epsilon^{CEM}_{\mathrm{trig}}=0.9656\pm0.0006$,
$\epsilon^{CMUP}_{\mathrm{trig}}=0.887\pm0.007$, and
$\epsilon^{CMX}_{\mathrm{trig}}=0.954\pm0.006$.  The factor
$k_{\mathrm{lep-id}}$ is evaluated by comparing a data sample of $Z$
+ jets events with a {\sc pythia} $Z$ sample, and found to be 1.00 for
electrons, 0.95 for CMUP muons, and 0.99 for CMX muons; because there
are few $Z$ + jets events at high jet multiplicity, we assign a 5\%
relative systematic uncertainty on $k_{\mathrm{lep-id}}$.

The efficiency $\mathcal{\epsilon}^\mathrm{MC}_{t\bar t}$ ~is
evaluated using a sample of {\sc pythia} $t\bar t$ Monte Carlo events
with top quark mass $m_t=175\,\mathrm{GeV}/c^2$.  Table~\ref{tab:acc}
summarizes the event selection acceptance for each type of lepton.
The $b$-tagging efficiency is measured from the same Monte Carlo
sample, and takes into account the $b$-tagging scale factor
(section~\ref{sec:secvtx}) by randomly keeping only 82\% of all the
tags, and discarding the others.  We find the efficiency for tagging
at least one jet in a $t\bar t$ event (after all other cuts have been
applied, including $H_T>200\,\mathrm{GeV}$) to be
53.4$\pm$0.3(stat.)$\pm$3.2(syst.)\%; the systematic uncertainty
comes from the measurement of the $b$-tagging scale factor, described
in Section~\ref{sec:secvtx}. 
An important source of uncertainty comes from the measurement of the jet energy, which
also affects the \met~and $H_T$ measurements (energy scale):
uncertainties relative to the $\eta$-dependent detector response,
overall energy scale, non-linearity, fraction of energy lost
outside the reconstructed jet cone, and multiple interactions
are added in quadrature.
Multiple interactions are soft interactions that can contribute 
to the jet energy measurement. They are not included in the simulation;
instead, a small average correction is applied to each jet in the data.
The energy of each jet is then shifted up and down in the MC by the
uncertainty, and half of the difference in the acceptance (4.9\%) is taken
as a systematic uncertainty.
The overall acceptance is 3.84$\pm$0.03(stat.)$\pm$0.40(syst.)\%, including all systematic
effects.  Table~\ref{tab:accsys} summarizes the dominant sources of
uncertainty for the acceptance.

\begin{table*}
\caption{\label{tab:acc} Summary table of the $t\bar t$ acceptance,
for a top quark mass of $175\,\mathrm{GeV}/c^2$.}
\begin{ruledtabular}
\begin{tabular}{lcccc}
       & CEM & CMUP & CMX & Total \\
\hline
Sample (total)      & 344,264 & 344,264 & 344,264 & 344,264 \\
\# Events w/o $b$-tag & 15,893 & 9791 & 3617 & 29301\\ 
Acc. w/o $b$-tag (\%) & 4.09$\pm$0.03$\pm$0.36 & 2.13$\pm$0.02$\pm$0.19 & 0.959$\pm$0.016$\pm$0.085 & 7.18$\pm$0.04$\pm$0.61 \\
\hline
\# Tagged Events    & 8490 & 5202 & 1965 & 15657\\
Tag Efficiency (\%) & 53.4$\pm$0.4$\pm$3.2 & 53.1$\pm$0.5$\pm$3.2 & 54.3$\pm$0.8$\pm$3.3 & 53.4$\pm$0.3$\pm$3.2\\
Acc. with $b$-tag (\%)& 2.19$\pm$0.02$\pm$0.23 & 1.14$\pm$0.01$\pm$0.12 & 0.512$\pm$0.009$\pm$0.054 & 3.84$\pm$0.03$\pm$0.40\\
Integ. Lumi. ($\mathrm{pb}^{-1}$) & 162$\pm$10 & 162$\pm$10 & 150$\pm$9 & \\ 
\end{tabular}
\end{ruledtabular}
\end{table*}

\begin{table}
\caption{\label{tab:accsys} Relative systematic uncertainties on the
  signal acceptance which are common to all lepton types.} 
\begin{ruledtabular}
\begin{tabular}{lc}
Quantity                   & Relative error (\%)    \\
\hline
$\epsilon_{z_0}$           & 0.5                    \\
Tracking Efficiency        & 0.4                    \\ 
Energy Scale               & 4.9                    \\
PDF                        & 2.0                      \\
ISR/FSR                    & 2.6                    \\
MC modelling               & 1.4                    \\
Lepton ID                  & 5.0                      \\
$b$-tagging                & 6.0                      \\
\end{tabular}
\end{ruledtabular}
\end{table}

%% file: single_tags.tex
\section{\label{sec:singletags}Cross-Section for Single-Tagged Events
  ($\ge$ 1 b-tags)}

\begin{figure}
  \includegraphics[width=0.5\textwidth]{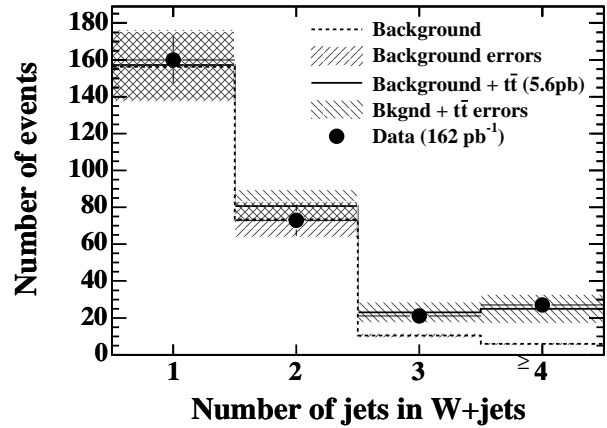}%
  \caption{\label{fig:njetstop}Background and $t\bar t$ signal
    expectation (based on measured $t\bar t$ cross section)
    as a function of jet multiplicity.  Events with three or more jets
  are required to have $H_T > 200\,\mathrm{GeV}$.}
\end{figure}

The production cross section follows from the acceptance measurement
and the background estimate:
\begin{equation}
\sigma_{t\bar t} = \frac{N_{\mathrm{obs}} - N_{\mathrm{bkg}}}{\epsilon_{t\bar
    t} \times \mathcal{L}},
\end{equation}
where $N_{\mathrm{obs}}$ and $N_{\mathrm{bkg}}$ are the number of total
observed and background events, respectively, in the $W$ + $\geq 3$ jet
bins (see Table~\ref{tab:corrbkgdsingle}); $\epsilon_{t \bar t}$ is the
signal acceptance (see Table~\ref{tab:acc}); and $\mathcal{L}$ is the
integrated luminosity.  Many of the predicted backgrounds are based on
the number of pretag data events, but that number includes a
significant contribution from $t\bar t$ ~events.  
After
subtracting this contribution from the pretag sample, the
dependent backgrounds are recalculated.  The final background
contributions for the single-tag selection are summarized in
Table~\ref{tab:corrbkgdsingle} and represented in Figure~\ref{fig:njetstop}.

\begin{table*}
\caption{\label{tab:corrbkgdsingle}Background summary for the
  single-tag selection.  The total backgrounds are given before and
  after the correction for $t\bar t$ events in the pretag $W$+jets sample.}
\begin{ruledtabular}
\begin{tabular}{ccc|cccc}
& & & \multicolumn{2}{c}{ $H_T >$ 0 GeV } & \multicolumn{2}{c}{ $H_T >$ 200 GeV } \\
Jet multiplicity & $W$ + 1 jet & $W$ + 2 jets & $W$ + 3 jets & $W$ + $\geq 4$
jets & $W$ + 3 jets & $W$ + $\geq 4$ jets  \\ \hline
Pretag & 15314 & 2448 & 387 & 107 & 179 & 91 \\
Mistags & $40.9 \pm 6.1$ & $17.0 \pm 2.4$ & $5.2 \pm 0.7$ & $ 2.6 \pm
0.4$ & $3.3 \pm 0.4$ & $2.3 \pm 0.3$ \\
$Wb\bar b$ & $37.0 \pm 11.2$ & $22.5 \pm 6.5$ & $5.0 \pm 1.3$ & $1.6
\pm 0.5$ & $2.8 \pm 0.8$ & $1.4 \pm 0.4$ \\
$Wc\bar c$ & $13.7 \pm 3.4$ & $8.0 \pm 2.2$ & $1.6 \pm 0.5$ & $0.6 \pm
0.2$ & $0.9 \pm 0.3$ & $0.5 \pm 0.2$ \\
$Wc$ & $34.5 \pm 9.0$ & $7.7 \pm 2.0$ & $1.4 \pm 0.4$ & $0.3 \pm 0.1$
& $0.7 \pm 0.2$ & $0.3 \pm 0.1$ \\
$WW$/$WZ$/$ZZ$,$Z\rightarrow \tau\tau$ & $2.2 \pm 0.4$ & $2.5 \pm 0.4$ &
$0.6 \pm 0.1$ & $0.1 \pm 0.0$ & $0.3 \pm 0.1$ & $0.1 \pm 0.0$ \\
non-$W$ QCD & $24.3 \pm 3.5$ & $10.5 \pm 1.9$ & $3.4 \pm 0.7$ & $1.4 \pm
0.4$ & $1.7 \pm 0.4$ & $1.2 \pm 0.3$ \\
single top & $2.6 \pm 0.3$ & $4.6 \pm 0.5$ & $1.1 \pm 0.1$ & $0.2 \pm
0.0$ & $0.8 \pm 0.1$ & $0.2 \pm 0.0$ \\
$Z$+HF & $1.1 \pm 0.3$ & $0.6 \pm 0.2$ & \multicolumn{2}{c}{$0.2 \pm
  0.1$} & \multicolumn{2}{c}{$0.10 \pm 0.05$} \\ \hline
Total & $156.3 \pm 19.1$ & $73.4 \pm 9.8$ & $18.5 \pm 2.2$ & $6.9 \pm
0.9$ & $10.5 \pm 1.3$ & $6.0 \pm 0.8$ \\ 
Corrected Total & $156.3 \pm 19.1$ & $73.4 \pm 9.8$ &
\multicolumn{2}{c}{$23.1 \pm 3.0$} & \multicolumn{2}{c}{$13.5 \pm 1.8$} \\
Data & 160 & 73 & 29 & 28 & 21 & 27 \\
\end{tabular}
\end{ruledtabular}
\end{table*}

\begin{figure} 
  \includegraphics[width=0.5\textwidth]{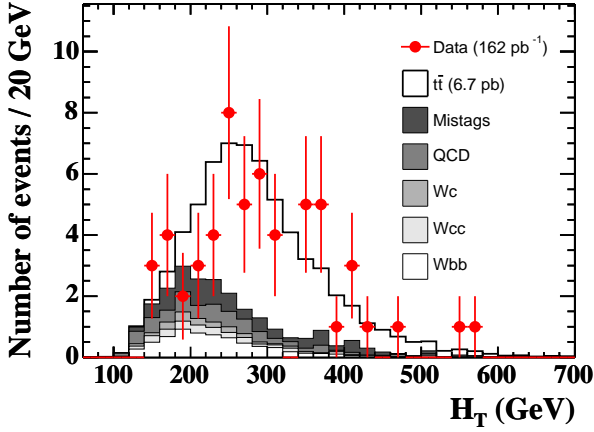}%
  \caption{\label{fig:htcand}$H_T$ distribution of the 57 tagged events with three or more jets, 
compared to the expected background and $t\bar t$ signal (normalized to the 
theoretical cross-section of 6.7 pb).} 
\end{figure}

\begin{figure}  
  \includegraphics[width=0.5\textwidth]{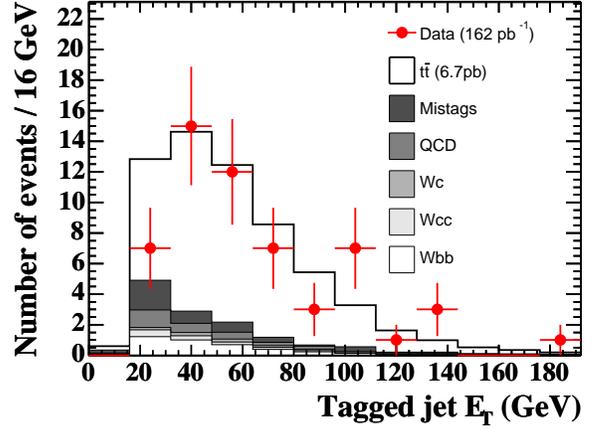}%
  \caption{\label{fig:etcand}$E_T$ distribution of the tagged jets in the 48 candidate events
with three or more jets and $H_T>200$~GeV, 
compared to the expected background and $t\bar t$ signal (normalized to the  
theoretical cross-section of 6.7 pb).}  
\end{figure}

\begin{figure}[h]   
  \includegraphics[width=0.5\textwidth]{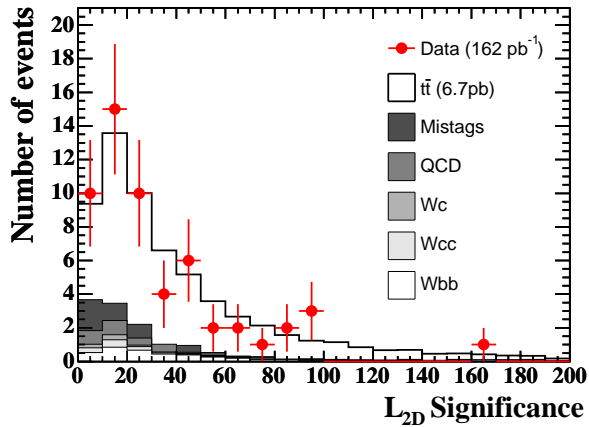}%

  \includegraphics[width=0.5\textwidth]{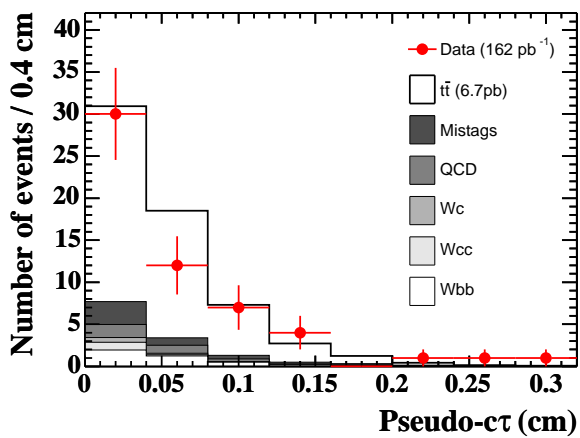}%
  \caption{\label{fig:ctaucand}Transverse decay length significance and pseudo-$c\tau$ distribution of the 
tagged jets in the 48 candidate events with three or more jets and $H_T>200$~GeV,   
compared to the expected background and $t\bar t$ signal (normalized to the   
theoretical cross-section of 6.7 pb).}   
\end{figure}

The properties of the selected candidate events are consistent with
the expectations for $t \bar t$ pair production and background
contributions.  Figures~\ref{fig:htcand} and ~\ref{fig:etcand} show
the distribution of the event $H_T$ and the tagged jet $E_T$, and
Figure~\ref{fig:ctaucand} shows the pseudo-$c\tau$ of the tagged jets.

For the optimized selection with the $H_T$ requirement, and for a top
quark mass $m_\mathrm{t} = 175\,\mathrm{GeV}/c^2$,
\begin{equation}
\sigma_{t\bar t} = 5.6^{+1.2}_{-1.1} \mathrm{(stat.)}
^{+0.9}_{-0.6} \mathrm{(syst.)} \mathrm{pb}.
\end{equation}
The systematic uncertainty is due to uncertainties on the signal
acceptance (10\% relative), luminosity measurement (6\%), and
background estimate (5\%). The acceptance, and therefore the measured
cross section, changes with the top quark mass as shown in
Table~\ref{tab:resultsvmass}.

\begin{table}
\caption{\label{tab:resultsvmass}Measured cross section for different
  top quark mass assumptions.}
\begin{ruledtabular}
\begin{tabular}{cc}
$m_t (\mathrm{GeV}/c^2)$ & $\sigma$ (pb) \\ \hline
170                      & $5.8^{+1.2}_{-1.1} \rm{(stat.)} ^{+0.9}_{-0.6} \rm{(syst.)}$ \\
175                      & $5.6^{+1.2}_{-1.1} \rm{(stat.)} ^{+0.9}_{-0.6} \rm{(syst.)}$ \\
180                      & $5.4^{+1.1}_{-1.0} \rm{(stat.)} ^{+0.9}_{-0.6} \rm{(syst.)}$ \\
\end{tabular}
\end{ruledtabular}
\end{table}

%% file: double_tags.tex
\section{\label{sec:dbtag}Cross Check using Double-Tagged Events
($\ge$ 2 $b$-tags)}

Each $t\bar t$ event contains two energetic $b$ quarks, making it
likely that two jets in the event will be tagged. Of the 57 tagged
events in the three and four jet bins before the $H_T$ cut, 8 of these
are double-tagged events.  The double-tag sample provides a cleaner
$t\bar t$ sample in which to cross-check the cross section with a
significantly smaller systematic uncertainty due to the background
estimate, although with decreased statistical precision.

\subsubsection{\label{sec:dbtagbk}Double-Tag Backgrounds}

The background estimate for the double-tag selection uses the methods
described in Section~\ref{sec:m2bkg}, except for a few additional
issues which are specific to the double-tag estimate.

The mistag estimate for double-tags is dominated by events with one
real tag of a heavy flavor jet with the second tag coming from the
mistag of an additional jet.  The mistag estimate is obtained by
applying the mistag matrix to the jets in the tagged sample, in
contrast to the pretag sample used for the inclusive estimate.  Since
the tagged sample with three or more jets is dominated by $t\bar t$
events, applying the mistag matrix to the entire tagged sample gives
an overestimate of the mistag background. Additionally, $Wb\bar b$ and
$Wc\bar c$ events with two heavy flavor jets are not counted as part
of the mistag estimate; rather, the mistag estimate is scaled by the
fraction of inclusive tagged events which are from mistags, $Wc$, and
non-$W$ QCD background.
 
The largest background comes from $Wb\bar b$ events, where both $b$ quark jets
are tagged. This background estimate uses the same heavy flavor
fractions and tagging efficiencies given in Section~\ref{sec:hffracs},
and is normalized to the pretag sample of $W$ + jet events. There is a
small additional contribution of double-tags in $Wb\bar b$ events where the
second tag is from mistags of light quark jets, so the mistag matrix
is applied to light quark jets in the Monte Carlo sample to account
for this additional contribution to the $Wb\bar b$ event double-tag
efficiency. The $Wc\bar c$ background is treated in the same way as
$Wb\bar b$.

The non-$W$ QCD background estimate uses the same lepton isolation and
$\met$ sideband regions described in Section~\ref{sec:qcd} to estimate
the double-tag background from direct production of heavy flavor jets.
There are zero double-tagged events with two or more jets in region B
($\cal I_{\rm sol}$ $<$ 0.1 and \met $<$ 15 GeV) compared to 133 single-tagged
events, implying a Poisson upper limit of 1.8\% at 90\% C.L. for a
single-tagged QCD event to be double-tagged.  This is applied to the
inclusive tag QCD background estimate given in
Table~\ref{tab:corrbkgdsingle}, and the limit is quoted as the
uncertainty on the background estimate of zero double-tagged QCD
events.

The double-tag backgrounds for the electroweak and single top processes
follow directly from the same Monte Carlo calculations discussed in 
Section~\ref{sec:mcbkg}. The only significant contributions come from
the s-channel single top process and $WZ$ with $Z\rightarrow b\bar b$.
The total double-tag background estimate is given in Table~\ref{tab:dbtag}.

\begin{table}
\caption{\label{tab:dbtag}Prediction for the number of double-tagged
events. Corrected total comes from the $t\bar t$ cross section
measurement where the pretag sample is corrected for the $t\bar t$
contribution. The expected number of $t\bar t$ events is calculated
using the measured cross section of 5.0 pb.}
\begin{ruledtabular}
\begin{tabular}{cc|cc}
Jet multiplicity & 2 jets & 3 jets & $\geq 4$ jets\\ 
\hline 
Single top  & 0.40$\pm$ 0.08 &  0.15$\pm$ 0.03 &  0.04$\pm$ 0.01 \\
 $WZ$     & 0.15$\pm$ 0.04 &  0.02$\pm$ 0.01 &  0.01$\pm$ 0.01 \\
 $Wb\bar b$     &    2.76$\pm$ 0.86 &  0.64$\pm$ 0.18 &  0.21$\pm$ 0.06  \\
 $Wc\bar c$     & 0.20$\pm$ 0.08 &  0.05$\pm$ 0.02 &  0.03$\pm$ 0.01 \\
 Mistag/QCD  & 0.14 $\pm$ 0.04 &  0.16 $\pm$ 0.04 &  0.11 $\pm$ 0.03 \\
 Total   &  3.65 $\pm$ 0.97 &  1.02 $\pm$ 0.23 &  0.40 $\pm$ 0.09  \\
\hline
Corrected Total &  3.6$\pm$ 1.0 & \multicolumn{2}{c}{1.3 $\pm$ 0.3 }  \\
\hline
$t\bar t$ ~(5.0 pb)   &  1.0   &  2.6   &  4.1 \\
 \hline
Data  &  8 &  3 &  5 \\
\end{tabular}
\end{ruledtabular}  
\end{table} 

\subsubsection{\label{sec:dbtagxsec}Double-Tag Acceptance and Cross Section}

For the double-tag analysis the backgrounds are sufficiently low that
we do not apply a cut on $H_T$. The pretag acceptance uses the same MC
sample and lepton identification and trigger efficiency corrections
described in Section~\ref{sec:singletags}.  The pretag efficiencies
are 4.32 $\pm$ 0.35\% for CEM electron, 2.24 $\pm$ 0.22\% for CMUP
muon and 1.01 $\pm$ 0.13\% for CMX muon $t\bar t$ events with three or
more jets.  The efficiency to double-tag $t\bar t$ events with three
or more jets is measured from Monte Carlo to be 0.11 $\pm$ 0.02 after
correcting for the difference in tagging efficiency between data and
Monte Carlo.

The total double-tag background estimate is given in
Table~\ref{tab:dbtag}.  The cross section for the double-tagged sample
is measured using the events in the three and four jet bins as for the
inclusive tagged sample, and correcting the pretag sample for the
$t\bar t$ contribution from double-tags.  Eight double-tag events are
observed on a background of 1.3 events, implying a cross section of
\begin{equation} \sigma_{t\bar t} = 5.0^{+2.4}_{-1.9}\mathrm{(stat)}
^{+1.1}_{-0.8} \mathrm{(syst)} \mathrm {pb}.  \end{equation}

The systematic error is due to the following contributions: tagging
efficiency~(15\%), acceptance~(7\%), luminosity~(6\%) and
backgrounds~(5\%). 

This result gives a consistent cross section measurement in an
almost background-free sample.  
With a larger data sample, this double-tag selection may
offer an improved measurement of the $t\bar t$ cross section.
In addition, the double-tagged sample may prove useful in estimating
relative contributions of the different $W$ + jets production
diagrams, especially gluon splitting to heavy flavor quark pairs.


%% file: conclusions.tex
%
\section{\label{sec:conclusions}Conclusions}

The $t\bar t$ production cross section has been measured with
vertex-tagged lepton + jets events from $162\,\mathrm{pb}^{-1}$ of
data collected at $\sqrt{s}=1.96\,\mathrm{TeV}$.  The selection yields
a sample of 48 candidate events with one lepton, large missing
transverse energy, large total transverse energy $H_T$, and three or
more jets, where at least one jet has a displaced secondary vertex
tag. A total of $13.5 \pm 1.8$ events are expected from non-$t\bar t$
processes.  The measured production cross section, assuming a top
quark mass of $175\,\mathrm{GeV}/c^2$, is
\[ \sigma(p\bar p \rightarrow t\bar t) =
5.6^{+1.2}_{-1.1} \mathrm{(stat.)}  ^{+0.9}_{-0.6} \mathrm{(syst.)} \ 
\mathrm{pb}.  \] 
Applying the same selection, except for the large transverse energy
requirement, to a double-tagged sample
(8 observed events with expected background of $1.3\pm
0.3$ events) yields a cross section of $5.0 ^{+2.4}_{-1.9}
\mathrm{(stat.)}  ^{+1.1}_{-0.8}\mathrm{(syst.)}\,\mathrm{pb}$.  Both
results are consistent with the theoretical predictions of $6.7
^{+0.7}_{-0.9}$ pb, again assuming
$m_t=175\,\mathrm{GeV}/c^2$~\cite{cacciari,kidonakis}.


%% file: secvtx_appendix.tex
%




\section{\label{app:HFeffect}Derivation of formulae for double-tag
  method of determining efficiency scale factor}
The measurement of the tagging efficiency in data employs the
double-tag method and uses identified conversions to estimate the
contribution of electrons which are fakes or part of a conversion
pair.  This appendix summarizes the detailed calculation of the
tagging efficiency.

Most of the electrons in the inclusive electron data sample (electron
$E_T > 8\,\mathrm{GeV}$ with no \met requirement) are from conversions or
fakes in light flavor jets. Away jet tagging enhances the heavy flavor
fraction in the electron side, but it still needs a significant light
flavor correction.  In general, the heavy flavor production in the
jets comes from three subprocesses: direct production, flavor
excitation and gluon splitting.  For simplicity, the final data sample
can be divided into the following four subclasses:

\begin{itemize} 
\item $N_{HH}$: the number of events where both sides contain a heavy 
  quark, either $c$ or $b$ (it includes the contributions of gluon
  splitting in both sides),
\item $N_{HL}$: the number of events where the electron side is heavy
  flavor and the away side is light flavor,
\item $N_{LH}$: the number of events where the electron is coming from fakes 
  and conversions and the away side contains heavy flavor,
\item $N_{LL}$: the number of events where both sides are light flavors. 
\end{itemize} 

By construction, we have 
\begin{displaymath}
  N_{HH}+N_{HL}+N_{LH} +N_{LL} =N
\end{displaymath} 
where $N$ is the 
total number of events passing the final selection. 
The heavy flavor contributions in the 
electron side can be determined using the measurement of heavy flavor fraction
(see Section~\ref{sec:secvtx}).
\begin{displaymath}
  N_{HH} + N_{HL} = F_{HF}\cdot N
\end{displaymath}

The $N_{LH}$ contribution can be determined
using the away tags in the conversion electron sample. 
Finally, the contribution of 
$N_{LL}$ is estimated using the mistags in the negative side. 

Let us use the following notation to help the derivation of 
efficiency measurement. 
\begin{itemize} 
\item $\epsilon_H'$: tagging efficiency of heavy flavor in the 
  electron jet,
\item $\epsilon_H$: tagging efficiency of heavy flavor in the away jet,
\item $\epsilon_L'$: mistag efficiency in the electron jets,
\item $\epsilon_L$: mistag efficiency in the away jets,
\item $N_{a+}$, $N_{a-}$, $N_{e+}$ and $N_{e-}$: are the number of positive,
  negative tags in the away jets and in the electron jets,
\item $N_{a+}^{e+}$, $N_{a+}^{e-}$, $N_{a-}^{e+}$ and $N_{a-}^{e-}$: are the 
  number of double-tags in the combination of positive or negative tags in 
  electron jet when the away tag is present, either positive or negative. 
\end{itemize} 

Applying the tag in the away jets, the numbers of positive and
negative tags are: 
\begin{equation}
  \epsilon_H \cdot N_{HH} + \epsilon_L \cdot N_{HL} + \epsilon_H \cdot 
  N_{LH} + \epsilon_L \cdot N_{LL} = N_{a+}
  \label{eq:Na+}
\end{equation} 
\begin{equation}
  \epsilon_L \cdot N_{HH} + \epsilon_L \cdot N_{HL} + \epsilon_L \cdot 
  N_{LH} + \epsilon_L \cdot N_{LL} = N_{a-}.
  \label{eq:Na-}
\end{equation}

By subtracting Equation~\ref{eq:Na-} from Equation~\ref{eq:Na+} , we get 
\begin{equation} 
  (\epsilon_H - \epsilon_L)\cdot (N_{HH}+N_{LH}) = N_{a+} - N_{a-}.
  \label{eq:stag_diff}
\end{equation} 

Applying the second tag on the electron side, the numbers of double-tags are 
\begin{equation} 
  \epsilon_H'\cdot (\epsilon_H-\epsilon_L) \cdot N_{HH} + 
  \epsilon_L' \cdot (\epsilon_H-\epsilon_L) \cdot N_{LH} =
  N_{a+}^{e+}-N_{a-}^{e+}
  \label{eq:dtag_diff_e+}
\end{equation} 

\begin{equation} 
  \epsilon_L'\cdot (\epsilon_H-\epsilon_L) \cdot N_{HH} + 
  \epsilon_L' \cdot (\epsilon_H-\epsilon_L) \cdot N_{LH} = N_{a+}^{e-}
  -N_{a-}^{e-}.
  \label{eq:dtag_diff_e-}
\end{equation}  

Subtracting Equation~\ref{eq:dtag_diff_e-} from
Equation~\ref{eq:dtag_diff_e+}, we get
\begin{equation} 
  (\epsilon_H' -\epsilon_L') \cdot (\epsilon_H - \epsilon_L) \cdot N_{HH} = 
  (N_{a+}^{e+} - N_{a+}^{e-}) - (N_{a-}^{e+} - N_{a-}^{e-}).
  \label{eq:dtag_diff}
\end{equation} 

From Equation~\ref{eq:stag_diff}, we get 
\begin{equation}
  (\epsilon_H - \epsilon_L)\cdot N_{HH} = (N_{a+}-N_{a-}) -
  (\epsilon_H-\epsilon_L)\cdot N_{LH}
  \label{eq:NBB_bgsub}
\end{equation}

Substituting Equation~\ref{eq:NBB_bgsub} into Equation~\ref{eq:dtag_diff} and rearranging terms, the heavy flavor tagging efficiency on the electron jet is 
\begin{displaymath}
  \epsilon_H' - \epsilon_L' = \frac{(N_{a+}^{e+} - N_{a+}^{e-}) - 
    (N_{a-}^{e+}-N_{a-}^{e-})}{ (N_{a+}- N_{a-}) - (\epsilon_H-\epsilon_L)\cdot
    N_{LH}}
\end{displaymath} 

In order to determine $N_{LH}$, we select the events where the electron is 
identified as a conversion partner and the away side is tagged. 
The heavy flavor contribution in the away jets should not depend on whether 
the electron originated from a photon conversion or a fake. 

Let us denote the following quantities 
\begin{itemize} 

\item $f$: the fraction of electrons originating from conversions in  
  no-heavy flavor jets in the electron side,
\item $f'$: the fraction of electrons originating from conversions in
  heavy flavor jets in the electron side,  
\item $\epsilon^c$: the efficiency of the conversion finding algorithm,
\item $\epsilon^o$: the error rate of finding a real electron as a part of 
  conversion, which is determined using the same sign,
\item $n^c$: the number of identified conversion electrons. 
\end{itemize} 

\begin{widetext}
Applying the conversion finding algorithm to the
data sample, the number of conversion electrons is
\begin{eqnarray}
  \label{eq:nc}
  (f'\cdot \epsilon^c + (1-f')\cdot \epsilon^o )\cdot (N_{HH}+N_{HL}) 
  + (f\cdot \epsilon^c + (1-f)\cdot \epsilon^o) \cdot (N_{LH}+N_{LL})
  = n^c \\
  f \cdot (\epsilon^c-\epsilon^o) \cdot (N_{LH} + N_{LL}) = 
  n^c - N \cdot \epsilon^o - f' \cdot (\epsilon^c - \epsilon^o) \cdot (N_{HH}+N_{HL}).
\end{eqnarray}  

By looking for conversions in the tagged electron jets, we have 
\begin{displaymath}
  (\epsilon_H' - \epsilon_L')\cdot (\epsilon^c \cdot f' + \epsilon^o \cdot (1-f')
  )\cdot (N_{HH} + N_{HL}) = n_{e+}^c - n_{e-}^c.
\end{displaymath} 

Since $(\epsilon'_H -\epsilon_L')\cdot (N_{HH} + N_{HL}) = N_{e+} -N_{e-}$
, we get 
\begin{equation}
  \label{eq:ec}
  \epsilon^o + f' \cdot (\epsilon^c - \epsilon^o) = \epsilon^{c'}
\end{equation} 
where $\epsilon^{c'} =  \frac{(n_{e+}^c - n_{e-}^c)}{ (N_{e+} - N_{e-})}$.

Substituting Equation~\ref{eq:ec} into Equation~\ref{eq:nc} , we have 
\begin{equation}
  \label{eq:fred}
  f \cdot (\epsilon^c - \epsilon^o) =
  \frac{n^c/N - (\epsilon^o + (\epsilon^{c'}  -\epsilon^o) \cdot F_{HF} )}{1-F_{HF}} 
\end{equation}

We apply the tag to the \emph{electron} jet in Equation~\ref{eq:nc} and the excess of tags is 
\begin{equation}
  \label{eq:excess_elec}
  f' \cdot (\epsilon^c - \epsilon^o) \cdot (\epsilon'_H - \epsilon'_L) \cdot 
  (N_{HH}+N_{HL})  = n^c_{e+} - n^c_{e-} - (N_{e+} - N_{e-}) \cdot \epsilon^{o}.
\end{equation}
If we then apply the tag to the \emph{away} jet in Equation~\ref{eq:nc} the excess of tags is 
\begin{equation}
  \label{eq:excess_away}
  f \cdot (\epsilon^c - \epsilon^o) \cdot (\epsilon_H - \epsilon_L) \cdot 
  N_{LH}  = n^c_{a+} - n^c_{a-} - (N_{a+} - N_{a-}) \cdot \epsilon^{o}
  - f'\cdot(\epsilon^c-\epsilon^o)\cdot(\epsilon_H-\epsilon_L)
  \cdot N_{HH}.
\end{equation}

Substituting Equations~\ref{eq:NBB_bgsub}~and~\ref{eq:ec}
into Equation~\ref{eq:excess_away}:
\begin{equation}
  \label{eq:barney}
  f\cdot(\epsilon^c-\epsilon^o)\cdot(\epsilon_H-\epsilon_L)\cdot
  N_{LH}
  = n^c_{a+} - n^c_{a-} - (N_{a+}-N_{a-})\cdot\epsilon^o - 
(\epsilon^{c'}-\epsilon^o)\cdot( (N_{a+}-N_{a-}) -  (\epsilon_H-\epsilon_L)\cdot N_{LH})
\end{equation}
\end{widetext}

From Equation~\ref{eq:fred} and Equation~\ref{eq:barney}, we get 
\begin{displaymath}
  (\epsilon_H- \epsilon_L) \cdot N_{LH} = (N_{a+}-N_{a-})\cdot 
  \frac{ \frac{n^c_{a+} - n^c_{a-}}{N_{a+} - N_{a-}} - \epsilon^{c'}}
  { n^c/N - \epsilon^{c'} }
  \cdot (1-F_{HF}).
\end{displaymath} 

Finally, the efficiency can be expressed as  
\begin{equation}
  \label{eq:eff_doubletag_full} 
  \epsilon_H' - \epsilon_L' = \frac{ (N_{a+}^{e+} -N_{a+}^{e-}) -(N_{a-}^{e+} - N_{a-}^{e-})} 
  {(N_{a+}-N_{a-})\cdot F^a_{HF}}  
\end{equation} 

\noindent where 
\begin{equation}
  \label{eq:bfrac_dtag}
  F^a_{HF} = 1 -   
  \frac{ \frac{n^c_{a+} - n^c_{a-}}{N_{a+} - N_{a-}} - \epsilon^{c'}}
  { n^c/N - \epsilon^{c'}}
  \cdot (1-F_{HF}).
\end{equation}